\documentclass[final,5p,times,twocolumn]{elsarticle}

\newcommand\BibTeX{{\rmfamily B\kern-.05em \textsc{i\kern-.025em b}\kern-.08em
T\kern-.1667em\lower.7ex\hbox{E}\kern-.125emX}}

\usepackage{amsmath,mathrsfs,amssymb,latexsym,epsfig} 
\usepackage{graphicx,psfrag,pstricks,verbatim,wasysym}
\usepackage{algorithm}
\usepackage{bclogo}
\usepackage[noend]{algpseudocode}
\usepackage[latin1]{inputenc}
\usepackage{subfig}
\usepackage{epsfig}
\usepackage{tabularx}
\usepackage{listings}
\usepackage{mathtools}
\usepackage{multirow}
\usepackage{tabu}
\usepackage{romanbar}
\usepackage{color}
\usepackage{subfig}
\usepackage{float}
\usepackage{placeins}

\newcommand{\I}{I}

\newcommand{\II}{\textit{II}}
\newcommand{\III}{\textit{III}}

\newcommand{\HO}{$\textit{HO} $ }
\newcommand{\HOns}{$\textit{HO} $}

\newcommand{\sym}{\text{sym}}
\newcommand{\C}{\bs{C}}

\newcommand{\F}{\bs{F}}
\newcommand{\Ev}[1]{\bs{e}_{\text{#1}}}
\newcommand{\ev}[1]{\bs{\varepsilon}_{\text{#1}}}
\newcommand{\bX}{\bs{X}}

\newcommand{\gradX}{\nabla_{\bf X}}

\newcommand{ \bs}[1]{\boldsymbol{#1}}

\newcommand{\PP}{\hbox{I\kern-.2em\hbox{P}}}
\newcommand{\HH}{\hbox{I\kern-.2em\hbox{H}}}

\newcommand{ \bC}{\boldsymbol{C}}
\newcommand{\bF}{\boldsymbol{F}}

\newcommand{\RR}{\hbox{I\kern-.2em\hbox{R}}}

\newcommand{\bu}{\bs{u}}
\newcommand{\bB}{\bs{B}}

\newcommand{\bx}{ \bs{x}  }

\DeclareMathOperator*{\argmin}{arg\,min}

\begin{document}
\title{A Viscoelastic Model for Human Myocardium}

\author[kcl,um]{David Nordsletten\corref{cor1}}
\author[kcl]{Adela Capilnasiu}
\author[um2]{Will Zhang}
\author[kcl]{Anna Wittgenstein}
\author[uc]{Myrianthi Hadjicharalambous}
\author[graz]{Gerhard Sommer}
\author[kcl,inserm]{Ralph Sinkus}
\author[graz,norway]{Gerhard A. Holzapfel}

\address[kcl]{Division of Biomedical Engineering and Imaging Sciences, Department of Biomedical Engineering, King's College London, UK}
\address[um]{Departments of Biomedical Engineering and Cardiac Surgery, University of Michigan, Ann Arbor, USA}
\address[um2]{Department of Biomedical Engineering, University of Michigan, Ann Arbor, USA}
\address[uc]{Department of Mechanical \& Manufacturing Engineering, University of Cyprus, Nicosia, CY}
\address[graz]{Institute of Biomechanics, Graz University of Technology, AT}
\address[inserm]{Inserm U1148, LVTS, University Paris Diderot, University Paris 13, Paris, FR}
\address[norway]{Department of Structural Engineering, Norwegian University of Science and Technology, Trondheim, NO}

\cortext[cor1]{North Campus Research Center, Building 20, 2800 Plymouth Rd, Ann Arbor 48109. \emph{e-}mail: nordslet@umich.edu}

\begin{abstract}
Understanding the biomechanics of the heart in health and disease plays an important role in the diagnosis and treatment of heart failure.
The use of computational biomechanical models for therapy assessment is paving the way for personalized treatment, and relies on accurate constitutive equations mapping strain to stress.
Current state-of-the art constitutive equations account for the nonlinear anisotropic stress-strain response of cardiac muscle using hyperelasticity theory.
While providing a solid foundation for understanding the biomechanics of heart tissue, most current laws neglect viscoelastic phenomena observed experimentally.
Utilizing experimental data from human myocardium and knowledge of the hierarchical structure of heart muscle, we present a fractional nonlinear anisotropic viscoelastic constitutive model.
The model is shown to replicate biaxial stretch, triaxial cyclic shear and triaxial stress relaxation experiments (mean error $ \sim 7.65 \%  $), showing improvements compared to its hyperelastic (mean error $ \sim 25 \%  $) counterparts.
Model sensitivity, fidelity and parameter uniqueness are demonstrated.
The model is also compared to rate-dependent biaxial stretch as well as different modes of biaxial stretch, illustrating extensibility of the model to a range of loading phenomena. 
\end{abstract}


\maketitle

\section{Introduction}

The biomechanical function of the human heart is a critical component of cardiac physiology. 
Beyond the role of its active properties leading to contraction of the myocardium, the passive characteristics of heart muscle play a key role in cardiac pathophysiology, particularly in conditions such as diastolic heart failure~\cite{zile2004diastolic}, heart failure with preserved ejection fraction (HFpEF)~\cite{sharma2014heart}, and myocardial infarction~\cite{cleutjens2002integration}.
Patients with these conditions often have poor outcomes due, in part, to patient variability and current challenges in predicting therapy efficacy.
Cardiac biomechanical modeling provides a tool for addressing these needs, providing the capacity for patient-specific assessment and model predicted outcomes.
These models are playing an increasingly important role in translational cardiac modeling~\cite{Chabiniok2016} and rely on appropriate constitutive models to predict the passive biomechanical response of the myocardium throughout the cardiac cycle.

Constitutive model characterization of passive myocardial tissue has been a focus of research for over 1.5 centuries~\cite{Woods:1892,blix1892lange}.
Experimental studies in animals have shown that myocardial tissue exhibits nonlinear stress-strain response~\cite{pinto1973mechanical}, anisotropy in biaxial stretch~\cite{demer1983,Yin1987,humphrey1990a}, and orthotropy under shear~\cite{dokos2000triaxial,dokos2002}.
Recently, these effects were extended and shown in bovine~\cite{avazmohammadi2018integrated,li2020insights} and human myocardial tissue~\cite{sommer2015quantification,sommer2015biomechanical}.
These experimental insights have driven the development of numerous mechanical constitutive models, with varying degrees of fidelity~\cite{Chabiniok2016}.
In most cases, myocardial models have leveraged hyperelasticity theory~\cite{holzapfel2000nonlinear, Bonet1997,truesdell2004non}, defining the stored energy (strain-energy) in response to the loading of muscle tissue.
Efforts at developing constitutive relations largely paralleled available data, with early descriptions focusing on transversely isotropic strain-energy equations~\cite{humphrey1989,humphrey1990a,humphrey1990b,guccione1991}, followed by orthotropic descriptions~\cite{costa2001modelling,Schmid2006,schmid2008,holzapfel2009}.
In the hyperelastic formulation, the transfer from external energy to internal energy (or \emph{vice versa}) is lossless, providing perfect energetic retention and return.

While current models treat the heart muscle as hyperelastic, this belies the considerable evidence of myocardial viscoelasticity.
Viscoelastic response has long been observed in muscle tissue, with early evidence stemming from the work of Blix~\cite{blix1892lange, blix1893lange, blix1894lange} who showed hysteresis in \emph{ex vivo} frog gastrocnemius.
This loss was further characterized by Hill and Hartree~\cite{hill1921thermo}, who demonstrated the loss of energy in the stretch and relaxation of viable/non-viable muscle tissue samples.
These \emph{viscous elastic} effects were initially explained using a \emph{spring and dashpot} model by Levin~\cite{levin1927viscous}.
Hysteresis and nonlinear stress-strain relations were later demonstrated in the canine papillary by Walker~\cite{walker1960potentiation} and at the organ-scale in the Langendorff feline heart experiments of Leach and Alexander~\cite{leach1965effect}. 
Viscoelastic relaxation phenomena were observed in the \emph{ex vivo} beating tortoise heart by O'Brien and Remington~\cite{obrien1966time}.
Similar experiments were studied in conscious dogs, demonstrating hysteresis and creep \emph{in vivo}~\cite{lewinter1979time}.
A comprehensive mechanical assessment was later performed in the rabbit papillary muscle by Pinto and Fung~\cite{pinto1973mechanical}, showing relaxation, creep, hysteresis along with a modest frequency dependence.
Since, viscoelastic behaviors have been reported in many experimental works~\cite{demer1983,Yin1987,dokos2000triaxial,dokos2002,sommer2015quantification,sommer2015biomechanical}.

The role of viscoelasticity in myocardial mechanics, while clear experimentally, has yet to be widely adopted the constitutive equations for heart tissue.
This is, in part, due to the already complex nature of state-of-the-art cardiac mechanics models -- involving multiple nonlinear anisotropic terms with multiple unknown parameters.
In addition, the lack of straightforward nonlinear viscoelastic models to build from has limited their extension.
A range of rheological analyses have been performed \cite{zhang2021compare}, and used as the basis for linear viscoelastic models (see, \emph{e.g.},~\cite{wilhelm1975viscoelasticity,fung2013biomechanics,dill2006continuum}).
These efforts were soon extended into nonlinear elasticity theory by Coleman and Noll~\cite{coleman1961foundations}, Truesdell and Noll~\cite{noll1958mathematical,truesdell2004non}, Pipkin and Rogers~\cite{pipkin1968non}, and extended to quasi-linear viscoelasticity for biological tissues by Fung~\cite{fung2013biomechanics} (see the review by Wineman~\cite{wineman2009nonlinear}).
Efforts moving nonlinear viscoelastic models into simulations were done by Simo~\cite{simo1987fully} and Holzapfel~\cite{holzapfel1996large,holzapfel1996new,holzapfel2001viscoelastic}.
Extension of these approaches to the nonlinear viscoelastic behavior in the heart were presented by Huyghe \emph{et al.}~\cite{huyghe1991} and Cansiz \emph{et al.} \cite{cansiz2015orthotropic}.
Recently, a study by Gultekin \emph{et al.}~\cite{gultekin2016orthotropic} used an analogue to the Maxwell-Wierchert model to characterize viscoelastic effects in the different microstructural orientations, providing the first three-dimensional anisotropic nonlinear viscoelastic constitutive equation for human myocardial tissues.
Interestingly, however, the model required significantly different parameter sets and relaxation times depending on the experiment performed by Sommer \emph{et al.} \cite{sommer2015quantification,sommer2015biomechanical}.

Here we present a three-dimensional viscoelastic constitutive model framework for the human myocardium.
A structural argument is presented based on the hierarchical nature of the myocardial tissue, suggesting the presence of a spectrum of relaxation times.
Following the previous works of Simo and others~\cite{simo1987fully,holzapfel1996large,holzapfel1996new} as well as extensions to fractional approximations~\cite{magin2006fractional,holm2019waves}, a fractional anisotropic nonlinear viscoelastic model is proposed that encapsulates phenomena observed in the heart.
The model developed for myocardium is fit to recent human myocardial data~\cite{sommer2015quantification,sommer2015biomechanical}, showing mean errors of $ \sim 7.65 \%$ compared to $ \sim 25 \% $ for hyperelastic variants.
Moreover, the model is used to predict variations of biaxial stretch and stretch rate, showing compelling predictions of the passive muscle response.
The developed model uses $ 11 $ parameters (compared to $ 17-18 $ used in~\cite{cansiz2015orthotropic,gultekin2016orthotropic}), which are shown to be unique.
This model represents the first nonlinear anisotropic viscoelastic model of human myocardium demonstrated to fit the biomechanical response of myocardial tissue and show reasonable predictions of tissue response.

Outlining the work, we begin by reviewing the potential sources of myocardial viscoelasticity, building a structural argument that provides the foundation of the model (section~\ref{sect:origins}).
Basic notation and hyperelastic formulations for cardiac mechanical models are reviewed (sections~\ref{sect:notation} and~\ref{sect:he}). 
From these foundations the anisotropic nonlinear viscoelastic power law model for the human myocardium is introduced (sections~\ref{sect:ve}-\ref{sect:proposed_vemodel}).
Sections~\ref{sect:data} and~\ref{sec:paramid} review the human rheological data used in this study as well as the parameterization procedure followed.
Results of the model are then presented in section~\ref{sect:results}, with discussion presented in section~\ref{sect:discussion}.

\begin{figure}[ht!]
\center
\includegraphics[width=\linewidth]{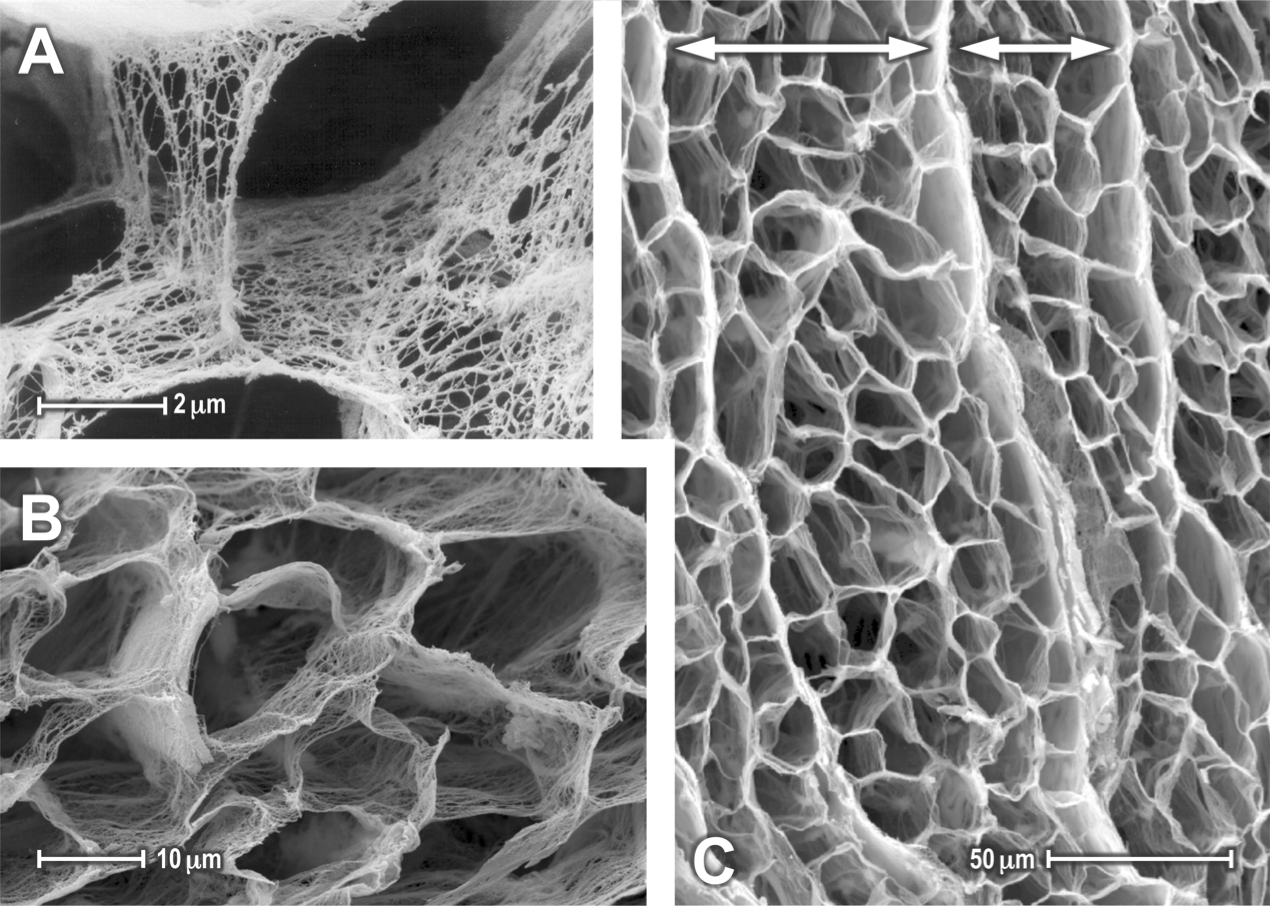}
\caption{Scanning electron microscopy of ECM structure in the (A) rabbit and (B,C) canine myocardium.  (A) Illustrates the detailed microstructure and collagen fibers composing the endomysial collagen layer that normally surrounds and interlinks cardiomyocytes.  (B) Shows magnification of the ECM structure, showing pockets normally encasing multiple myocytes as well as the coronary microvasculature.  (C) Magnification of the ECM, showing myocardial sheets separated by perimysial collagen layers (marked with arrows).  Reproduced with permission from~\cite{Macchiarelli2002,benedicto2011structural}.}\label{F:ECM}
\end{figure}

\section{Materials and Methods}~\label{sect:matmeth}
\vspace{-10mm}

\subsection{Origins of viscoelasticity in passive myocardium} \label{sect:origins}

The complex structure of myocardial tissue has led to discussion over the origins of viscoelasticity.
Many constituent components of the myocardium have been implicated as the source of viscoelasticity, including tissue perfusion, extracellular fluid, myocytes, the extracellular matrix (ECM) and others.
Simulated poromechanical studies~\cite{yang1991} have demonstrated that tissue perfusion can yield stress relaxation.
Increased extracellular fluid content is known to significantly influence the biomechanics of tissues~\cite{cavalcante2005mechanical,guo2007effect,azeloglu2008heterogeneous}.
Experimental studies on sarcomeres have demonstrated that the primary contractile proteins of the heart exhibit passive viscoelastic behavior~\cite{de1992internal}.
Studies have also shown that the main constituents of the ECM exhibit viscoelasticity~\cite{shen2011viscoelastic}, suggesting molecular friction as a source of viscoelastic response.
While these factors are often considered and advocated for individually, it is highly likely that all factors can contribute to the viscoelastic response of the myocardium with varying degrees of importance depending on the spatiotemporal scales and loading conditions considered.
In the following section, we review the evidence for these different factors contributing to the viscoelastic response of myocardial tissue.

\subsubsection{Influence of tissue perfusion and extracellular fluid}

Due to the surrounding interstitial fluid space and the perfusion of myocardial tissues by the coronary vasculature, it is clear that the heart is a complex poromechanical organ. 
Debate then arises whether the viscoelastic behavior observed predominantly stems from fluid movement through the porous tissue (\emph{i.e.} the tissue is nearly poroelastic) or if the solid compartment itself is viscoelastic (\emph{i.e.} the tissue is poroviscoelastic).
Questions also arise regarding the influence \emph{in vivo} versus the typical testing occurring \emph{ex vivo}.
In tissue studies by ~\cite{dokos2000triaxial,dokos2002,sommer2015biomechanical,avazmohammadi2018integrated}, non-perfused tissues exhibited significant viscoelastic response.
While this could be explained by the movement of interstital fluid, the shear rates required to dissipate the energy observed experimentally would require much larger frequencies and would not explain the seemingly nonlinear loss response observed.
This was shown through modeling by Yang~\cite{yang1991}, who demonstrated that the viscoelastic response due to poroelasticity was not sufficient to explain hysteresis observed in data.
However, the presence of extracellular fluid and vasculature has a clear influence on the biomechanics of tissue, with results showing that a change in the aqueous solution directly impacts apparent stiffness~\cite{meghezi2012effects}.
Hydration of myocytes and the ECM proteins both have significant impacts on their properties and viscoelasticity, making these factors critical to the passive behavior of tissue.
Further, in simulation studies~\cite{reeve2014constitutive}, it was shown that pore pressure yields a general stiffening by loading the ECM which could, in turn, influence viscoelastic response of structural proteins.
As a consequence, the viscoelastic response of tissue is inextricably linked to the constituents of the extracellular environment whether (or not) the porous flow of fluid plays a leading role in the viscoelastic mechanical response observed.

\subsubsection{Influence of cardiomyocytes}

The passive mechanics of cardiomyocytes have also been identified as contributors to viscoelastic response.
The viscoelastic impact of myocytes is often thought to be significant due to the encapsulated fluid of the cytosol.
However, skinned myocytes exhibit strong passive viscoelastic response, thought to arise from molecular friction in the macromolecules such as titin~\cite{hoskins2010normal}. 
In addition, intracellular structural proteins, such as actin~\cite{lieleg2010structure}, have also exhibited viscoelasticity due to molecular friction.
While myocytes exhibit passive viscoelastic response, it is unclear the degree to which passive cellular forces contribute to the total tissue response.
Witzenburg \emph{et al.}~\cite{witzenburg2012mechanical} demonstrated that decellularization of rat myocardium increased the apparent stiffness of the myocardium 6.7 fold in biaxial tests; an increase that corresponded to the 5.6 fold reduction in cross-sectional area.
While this may suggest that cells provide a minor contribution to passive biomechanics, it is clear that the turgid cylinder-like structure of the cardiomyocyte plays a significant role in resistance to shear and compression.

\begin{figure}[ht!]
\center
\includegraphics[width=\linewidth]{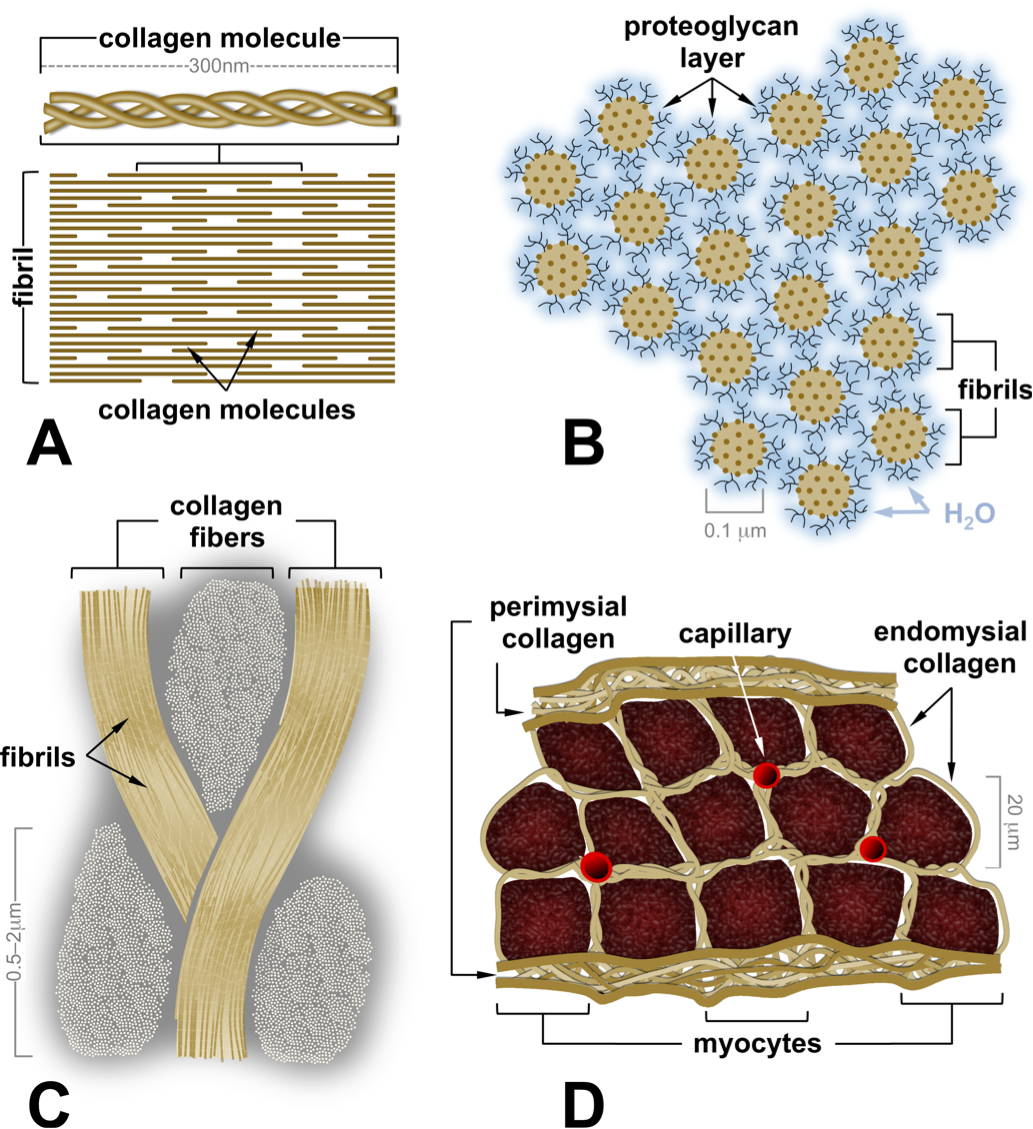}

\vspace{5mm} 

\includegraphics[width=\linewidth]{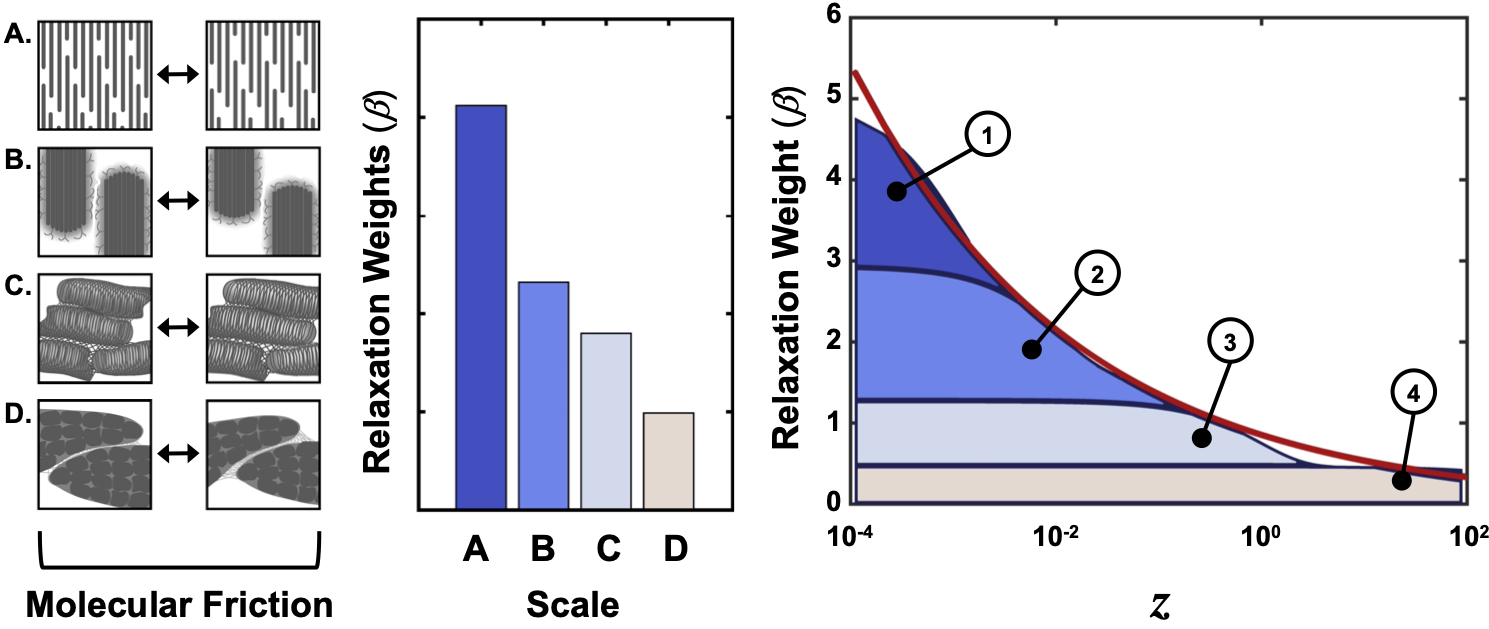}
\caption{
Schematic representation of the hierarchical structure of the extracellular matrix, and its representation as a power spectrum.  
(A) Illustration of the collagen triple helix and structural arrangement of the collagen fibril.
(B) Illustration of the lattice arrangement of fibrils within a collagen fiber, showing fibrils, proteoglycans and their hydration.
(C) Illustration of bundled fibrils within collagen fibers and their cross section.
(D) Illustration of the myocardial tissue showing myocytes and capillaries surrounded by endomysial collagen fibers and encased in sheets covered by perimysial collagen.
(Bottom Row) Illustration of multiscale friction processes yielding a fractional relaxation response moduli.  (Left) Sketch of multiscale friction processes (A) within fibrils, (B) between fibrils, (C) at the endomysial collagen scale, and (D) at the perimysial collagen scale.  
(Middle) Effective density of relaxation phenomena within a representative volume.  
(Right) Combined multiscale relaxation response modulus with colors indicating different scales of response and its representation using a fractional model (red curve).} \label{F:PowerLawArgument} 
\end{figure}

\subsubsection{Influence of extracellular matrix structure}

The extracellular matrix is acknowledged as a critical, and dominant, component of passive mechanical properties in the heart~\cite{weis2000myocardial,legrice2005architecture,lunkenheimer2006myocardium}.
The ECM is highly hierarchical~\cite{fratzl2008collagen} (see figure~\ref{F:ECM}), exhibiting important structural organization at multiple scales.
A fundamental building block of myocardial ECM are collagen I and III~\cite{bishop1999regulation} (approximately $300$ nm in length).  
The extremely rigid collagen molecule triple helices are bound to form collagen fibrils, with diameters of $10-500 $ nm and lengths on the order of $10-30$ $\mu$m (see figure~\ref{F:PowerLawArgument}A).
This structural arrangement provides opportunity for significant molecular interaction, yielding potential molecular friction mechanisms as molecules translate relative to one another.
Indeed, Shen \emph{et al.}~\cite{shen2011viscoelastic} performed stress relaxation experiments on individual collagen (type I) fibrils, demonstrating viscoelastic response.
Characterization of this relaxation response required a 2-component Maxwell-Weichert model, suggesting multiple relaxation times are present at the scale of individual fibrils.
It has also been shown that collagen fibrils exhibit nonlinear response, stemming from an uncoiling of end groups leading to progressive recruitment of molecules~\cite{fratzl1998fibrillar,puxkandl2002viscoelastic}.

Collagen fiber formation bundles together many individual fibrils.  
Fiber bundles vary significantly in diameter (with type I forming larger bundles than type III) and can span over extended distances.
To avoid rigid locking between fibrils, a layer of proteoglycans and glycosaminoglycans cover the outer fibril surface~\cite{hayes2013effect} (see figure~\ref{F:PowerLawArgument}B).
These large proteins are strongly hydrophilic, binding with water and enabling molecular lubrication between fibrils.
Collagen fibers are then linked together to other collagen molecules, often running in close proximity through myocardial tissue (see figure~\ref{F:PowerLawArgument}C).

At the tissue scale, fibers weave together forming the ECM (see figure~\ref{F:ECM}).
Fibers form complex layers of endomysial collagen -- surrounding and spanning between individual myocytes -- and perimysial collagen -- surrounding and spanning between muscle sheets (see figure~\ref{F:PowerLawArgument}D).
Endomysial and perimysial collagen exhibit a complex structure that undergoes predictable molecular realignment and uncoiling under load~\cite{purslow1989strain,hanley19993,purslow2008extracellular}.
Stretch of individual cells requires deformation of the endomysial layer, resulting in molecular friction as fibers translate relative to one another.
Scaling up to the level of myocardial sheets, a similar process of deformation and relative translation occurs within the perimysial layer.

Examining the hierarchical structure of the ECM, it is clear that molecular friction and potential drivers of viscoelastic response are pervasive and present at multiple spatial scales.
From fibril to fiber to collagen layers, a relative translation of molecules to macromolecular complexes can be observed.
The multiscale mechanisms of molecular friction suggest that myocardial tissue is likely characterized by a spectrum of relaxation events occuring at a broad range of time-scales.
This is in contrast to some collagen hydrogels which can exhibit less orderly structure~\cite{mori2012dynamic,mori2013dynamic} and can be well characterized by a simple series of relaxation times~\cite{li2015modeling, xu2013understanding,xu2013experimental,irastorza2015mathematical}.

\subsection{Kinematics and notation}\label{sect:notation}

Here, we briefly introduce the classic kinematic notation used in nonlinear mechanics (see, \emph{e.g.},~\cite{Bonet1997,ogden1997non,holzapfel2000nonlinear}).
Deformation in the heart can be defined by the motion of material points as they move from their reference, $\Omega_0 \subset \mathbb R^3 $, to their physical configurations, $ \Omega_t  \subset \mathbb R^3 $ (at some time $ t \in [0,T] $).
Marking material points of the reference domain by their coordinate position, $ \bX \in \Omega_0 $, the relative motion is defined by the displacement field $ \bu : \Omega_0 \times [0,T] \to \mathbb{R}^3 $ whereby $ \bx(\bX, t) = \bu(\bX,t) + \bX $.
Local deformation of material axes and volumes is characterized by the deformation gradient, $ \bF = \gradX \bu + \bs{I} $, and its determinant, $ J = \det \bF > 0 $, respectively~\cite{holzapfel2000nonlinear}.
Often, the heart is approximated as an incompressible tissue (\emph{e.g.}, $ J = 1 $), though debate about this exists in the literature~\cite{avazmohammadi2020vivo}. 
Measures of stretch are given by the right and left Cauchy Green tensors, defined as $ \bC = \bF^T \bF $ and $ \bB = \bF \bF^T $, respectively.

Often it is convenient to define constitutive relations using invariants of stretch / strain tensors (here generically denoted by $ \bs A $) given by~\cite{Bonet1997}
\begin{equation}
\I_{\bs A} = \bs A : \bs{I}, \quad \II_{\bs A} = \bs A : \bs A, \quad \III_{\bs A} = \det \bs A .
\label{eq_invariants}
\end{equation}
It can also be convenient to consider isochoric invariants, replacing $ \bs A $ with $ \bs{ \bar {A}} = \bs A / \III_{\bs A}^{1/3} $ in Eq.~\eqref{eq_invariants} (\emph{e.g.}, $ \bar{\I}_{\bs{ A}} = \bs{ \bar {A}} : \bs{I} $), as these can facilitate the isochoric / deviatoric split in hyperelastic formulations.

Considering anisotropic materials, such as myocardium, a common approach is to define mutually orthogonal local microstructural directions, \emph{e.g.}, $ \{ \Ev{f}, \Ev{s}, \Ev{n} \} $, at each material point. 
Here $ \Ev{a} $ denotes the unit orientation vector along myofibers (f), sheets (s) and sheet normals (n)~\cite{legrice1995}.
The physical orientation of these microstructural directions is given by $ \ev{a} = \bs{F} \Ev{a} $, $ a \in  \{ \text{f,s,n} \}  $.
Hence, the squared stretch along microstructural directions can be expressed by the pseudo invariants,
\begin{equation}
\I_{a} = \C : \Ev{a} \otimes \Ev{a} = \Ev{a} \cdot (\C \Ev{a}) = \ev{a} \cdot \ev{a}, 
\quad a \in \{ \text{f,s,n} \}  .
\label{eq_aniso_inv1}
\end{equation}
Generalizing Eq.~\eqref{eq_aniso_inv1}, 
\begin{equation}
\I_{ab} = \C : \sym ( \Ev{a} \otimes \Ev{b}) = \ev{a} \cdot \ev{b},
\quad a,b \in \{ \text{f,s,n} \} ,
\label{eq_aniso_inv}
\end{equation}
where $ \sym (\bs{A}) = \frac{1}{2} (\bs{A} + \bs{A}^T) $ is the symmetric transformation.  
These invariants play an important role in defining the constitutive behavior in myocardial tissue~\cite{holzapfel2009}. 

\subsection{Hyperelastic formulations of the myocardium}\label{sect:he}

Though a number of constitutive equations have been developed to describe myocardial tissue~\cite{Schmid2006,schmid2008,holzapfel2009}, in this study we focus on two of the most common, the Holzapfel and Ogden~\cite{holzapfel2009} (\HOns) and the Costa~\cite{costa2001modelling} models.
The \HO model is a hyperelastic incompressible strain-energy function, widely employed in modeling studies and is appealing for both theoretical reasons (\emph{e.g.}, convexity and objectivity) as well as parameter identifiability~\cite{schmid2008,Hadjicharalambous2015}.
Recalling briefly, the incompressible \HO strain-energy $ W_e : \mathbb{R}^{3\times 3} \times \mathbb{R} \to \mathbb{R}_{\ge0} $ can be written as
\begin{equation}
    W_e(\C, p) 
    = 
    W_{iso} ( \bar{\I}_{\C} ) 
    + 
    W_\text{ff} (\I_\text{ff}) 
    +
    W_\text{ss} (\I_\text{ss}) 
    +
    W_\text{fs} (\I_\text{fs}) 
    + 
    p(J-1),
    \label{eq_hyperelastic_se_ho}
\end{equation}
where
\begin{align}
    W_{iso} (\bar{\I}_{\C}) 
    & = 
    \frac{\text{a}}{2\text{b}} \big[ \exp \{ \text{b} ( \bar{\I}_{\C} - 3) \} - 1 \big], 
    \label{eq_hyperelastic_se_wiso} \\
    W_\text{kl} (\I_\text{kl} ) 
    & = 
    \frac{\text{a}_\text{kl}}{2 \text{b}_\text{kl}}  \big [ \exp \{ \text{b}_\text{kl} ( \I_\text{kl} -  \delta_\text{kl})^2 \}  -1 \big],
    \label{eq_hyperelastic_se_waniso}
\end{align}
and $ \delta_{kl} $ denotes the Kronecker delta (zero unless $ k = l $, in which case its $1 $).
In this formulation, the strain-energy is written as a sum of  components.  
The isotropic term, $W_{iso} $, provides the isotropic bulk distortional energy associated with tissue deformation.
Anisotropic terms $ W_\text{kl} $ (where $ \text{k,l} \in   \{ \text{f,s,n} \} $) are introduced to account for the varying distortional energy associated with deformation along microstructural directions.  
We note that the anisotropic invariants $ \I_\text{ff} $ and $ \I_\text{ss} $ are thought to not support compression, with $W_\text{ff}$ and $W_\text{ss} $ set to zero when $ \I_\text{ff}, \I_\text{ss} < 1 $~\cite{holzapfel2009}.
In the \HO model, the set of anisotropic terms depend on $ 8 $ fitting parameters (denoted by $ a $'s and $ b $'s).
In the \HO model, the second Piola Kirchhoff stress (PK2) $ \bs{S} $ can be written as
\begin{equation}
    \bs{S}  
    =  
    \frac{1}{ J^{2/3} } \frac{\partial W_{iso}}{\partial \bar{\I}_{\C} } \, \left ( \bs{I}  - \frac{\I_{\C}}{3} \C^{-1} \right )
    +  
    \sum_{\text{kl} \in S } \frac{\partial W_\text{kl}}{\partial \I_\text{kl} }  \,  \text{sym} ( \Ev{k} \otimes \Ev{l} )
    + \bs{S}_p,
\label{eq_pk2_ho} 
\end{equation}
with $ S =\{ \text{ff}, \text{ss}, \text{fs} \} $ and $ \bs{S}_p = pJ \C^{-1} $.
Through push forward operations (\emph{i.e.}, $ \bs{\sigma} = J^{-1} \bF \bs{S} \bF^T $), the second Piola Kirchhoff stress tensor can be extended into the Cauchy stress tensor.
We note that the original \HO formulation considered the strain-energy defined using standard invariants~\cite{holzapfel2009}, with later forms defined using isochoric invariants and/or dispersion~\cite{gultekin2016orthotropic}.

Another constitutive equation that is widely employed is the Costa model \cite{costa2001modelling}. 
This model is well-posed in terms of convexity and objectivity \cite{holzapfel2009} and relies on an orthotropic formulation of the exponential Fung-type law. 
While typically written in terms of rotated Green Lagrange strain tensors, the Costa model can similarly be expressed in terms of invariants of the right Cauchy Green strain tensor, \emph{e.g.}
\begin{equation}
    W = \frac{C}{4} \left [ 
        \mathcal{W}(\C)
        -
        1
    \right ] 
    + 
    p(J-1),
    \quad
    \mathcal{W}(\C) = \prod_{\text{kl} \in S^\prime} W_\text{kl}^\prime ( \I_\text{kl}),
    \label{eq_costalaw}
\end{equation}
where
\begin{equation}
    W_\text{kl}^\prime ( \I_\text{kl})
    = 
    \exp \{ \text{b}_\text{kl} ( \I_\text{kl} -  \delta_\text{kl})^2 \},
    \label{eq_hyperelastic_costaw} \\
\end{equation}
and $ S^\prime =\{ \text{ff}, \text{ss}, \text{nn}, \text{fs}, \text{fn}, \text{sn} \} $. 
This form is useful for comparison of these models, illustrating that the Costa model shares the same underlying exponential invariant formulation; however, strain-energy terms are grouped in one term $ \mathcal{W}(\C) $, through multiplication.
In this case, the Costa model is comprised of $ 7 $ material parameters, an outer scaling constant, $ C $, along with anisotropic scaling constants, b.
The PK2 stress tensor can be written as
\begin{equation}
    \bs{S} 
    =  
    C \mathcal{W}(\C)
    \sum_{\text{kl} \in S^\prime } \text{b}_\text{kl} ( \I_\text{kl} -  \delta_\text{kl})
    \text{sym} ( \bs{e}_k \otimes \bs{e}_l )
    +
    \bs{S}_p.
\label{eq_pk2_costa} 
\end{equation}

Both \HO and Costa models utilize exponential forms commonly employed in collagenous biological tissues~\cite{fratzl1998fibrillar,puxkandl2002viscoelastic} reflecting the toe, heel and linear regime of collagen fibrils and the progressive recruitment of coiled fibers with stretch.  
Further, while in this paper we focus on incompressible forms, both models have been used considering the myocardium as compressible (or nearly incompressible), where $ \bs{S}_p $ is replaced with an appropriate compressibility term.
The primary variation between models is the separable form employed in \HO compared to the coupled exponential form of the Costa model. Both model forms have advantages and disadvantages in terms of model performance and parameter identifiability.

\subsection{Extension to cardiac viscoelasticity}\label{sect:ve}

Viscoelastic phenomena in muscles have been described using a range of linear spring, dashpot and spring pot system models \cite{levin1927viscous,mainardi2010fractional}.  
One of the simplest models that captures stress relaxation (impulse stretch) and creep (impulse stress) is the Zener model, containing a Maxwell model in parallel with a spring (cf.~\cite{holzapfel2000nonlinear,dill2006continuum}).  
Denoting the stress in the elastic branch of the model as $ \sigma_0^e (t) = E_0 \varepsilon(t) $, the total stress in the Zener model can be written as
\begin{equation}
\sigma(t) =  \sigma_0^e (t) + \int_0^t B \exp \{ (s-t) / \tau \} \dot{\sigma}_0^e (s) \; d s
\label{eq_linear_zener}
\end{equation}
where $ \tau = (E / \eta)^{-1} $ is the relaxation time and $ B = E / E_0 $ is the relaxation weight~\cite{zhang2021compare}.  
In this model, the asymptotic elastic response is given by the first term, while the viscoelasticity is encapsulated in the second term.
Eq.~\eqref{eq_linear_zener} provides a template for defining the stress response \cite{adolfsson2003fractional}.  
Extending this idea to nonlinear materials, a straightforward approach is to replace the elastic stress component $ \sigma_0^e $ with its nonlinear hyperelastic variants.  
Replacing $ \sigma_0^e $ with an appropriate nonlinear PK2 form $ \bs{S}_e $, yields the quasi-linear viscoelasticity (QLV) Fung model,
\begin{equation}
    \bs{S}(t) 
    =  
    \bs{S}_e (t) 
    + 
    \int_0^t \mathcal{K} (s-t) \dot{\bs{S}}_e (s) \; d s 
    + 
    \bs{S}_p (t),
    \label{eq_pk2_1relaxation}
\end{equation}
where, for the Zener model, $ \mathcal{K} (s-t) = B \exp \{ (s-t) / \tau \} $.
Here the QLV Fung model extension is considered linear due to the linear dependence on $ \bs{S}_e $.
As noted in section~\ref{sect:origins}, myocardial tissue is strongly influenced by viscoelastic phenomena across a hierarchy of scales, with collagen fibrils themselves exhibiting multiple relaxation times~\cite{shen2011viscoelastic}.
A straightforward adaptation is to consider a nonlinear variant of the Maxwell-Wiechert model (or Generalized Maxwell model)~\cite{tschoegl1989phenom,schiessel1995generalized}.  
Considering $n$ viscoelastic elements in parallel, due to linearity, this could be modeled simply by altering the QLV form, with the relaxation 
\begin{equation}
    \mathcal{K} (s-t) 
    =  
    \sum_{k=0}^{n} B_k \exp \{ (s-t) / \tau_k \}.
    \label{eq_RRM_disc}
\end{equation}
Here, $ B_k $ and $ \tau_k $ represent the relaxation weights and relaxation times of the various viscoelastic elements.
Note that, here, we have assumed that the branches share the same nonlinear form $ \bs{S}_e $, which need not be the case.
While this form provides generality for encapsulating the viscoelastic response of a material in the QLV framework, it also requires unique determination of all the relaxation weights and times, which is challenging to obtain experimentally particularly for nonlinear anisotropic materials like the myocardium.
However, if these weights and time constants were distributed in a well-characterized way, it's possible that the spectrum could follow a simplified approximate form.

In myocardial tissue, viscoelastic response has been observed stemming from molecular friction processes, flow or intracellular effects (see section~\ref{sect:origins}), all of which occur at one or more spatiotemporal scales.
In the context of myocardial ECM, friction processes are observed at the basic scale of the fibril and extend, hierarchically, as we consider different scales within the ECM (see figure~\ref{F:PowerLawArgument}).
To argue for a specific form, we consider a representative volume of myocardial tissue.
At the microscale, the relaxation times are relatively small reflecting the fact that friction occurs between molecules at small spatial scales.
Bridging toward larger length scales, the relaxation times increase reflecting that friction occurs between larger conglomerates of molecules or whole tissue structures.
The effective density of these different mechanisms also varies across scales.
Within a representative volume, the occurrence of microscale phenomena is abundant, while the occurrence decreases toward larger spatial scales.
Under these heuristics arguments, the composite relaxation weight could be envisioned as a power law distribution, \emph{i.e.} $ \mathcal{K}(z)  \propto z^{-\alpha} $ (see figure~\ref{F:PowerLawArgument}).

\subsection{Extension to fractional viscoelasticity}

Based on the argument that, due to the hierarichal structure of myocardial tissue, the relaxation function $ \mathcal{K} $ resembles a power distribution, the QLV model simplifes to the nonlinear fractional Kelvin-Voigt model, \emph{i.e.}
\begin{equation}
    \bs{S}(t) = \bs{S}_e (t) + B_0 D^\alpha_t \bs{S}_e,
    \label{eq_pk2_springpot_general}
\end{equation}
where $ D_t^\alpha $ denotes the $\alpha$-order Caputo derivative~\cite{podlubny1998fractional}
\begin{equation}
    D^\alpha_t g  
    = 
    \frac{1}{\Gamma(1-\alpha)} \int_0^t (t - s)^{-\alpha} \, \dot{g}(s) \, ds, \quad \alpha \in [0,1]
    \label{eq_caputo}
\end{equation}
and $ B_0  $ is a strictly positive constant.
Here, $ B_0 $ modulates the relative strength of the viscoelastic response, while $ \alpha $ modulates the distribution of relaxation times.  
Indeed, for $ \alpha = 0$ the relaxation spectrum becomes infinitely dependent on instantaneous time-scales, reducing the term to a pure elastic response.
In contrast, as $ \alpha $ tends to 1, then the fractional operator becomes a nonlinear dashpot.
Use of this model significantly reduces the complexity and number of parameters required to construct the material response.  
This simplification is not for free; however, but instead reflects an inbuilt assumption that the relaxation spectrum exhibits power-law behavior.

Generalizing this framework, the viscoelastic PK2 stress tensor can be conveniently written as a fractional viscoelastic differential equation, \emph{e.g.}
\begin{equation}
    \bs{S}(t)  + \delta D_t^{\alpha_\delta} \bs{S} = \bs{S}_{e} (t) + D^\alpha_t \bs{S}_v,
    \label{eq_fracdifeq}
\end{equation}
or the analogous form,
\begin{equation}
    \bs{S}(t) 
    = 
    \bs{S}_{e} (t) + \bs{S}^\star (t),
    \quad
    \bs{S}^\star (t)+ \delta D_t^{\alpha_\delta} \bs{S}^\star 
    =  
    D^\alpha_t \bs{S}_v,
    \label{eq_fracdifeq2}
\end{equation}
where $ \bs{S}_v$ and $ \bs{S}_{e} $ denote the nonlinear viscoelastic and elastic responses, and $ \delta > 0 $ is a scaling weighting the history dependence on the PK2 stress tensor itself.
Note, this is a departure from the linearity seen in the QLV model.
Eq.~\eqref{eq_fracdifeq} was shown to exhibit realistic viscoelastic behavior across a range of testing scenarios \cite{zhang2021compare}, providing advantages to the fractional form in Eq.~\eqref{eq_pk2_springpot_general}.
Eq.~\eqref{eq_fracdifeq2} performs similar to Eq.~\eqref{eq_fracdifeq}, with the primary difference that the model retains a purely elastic term $ \bs{S}_e $.

\subsection{A fractional viscoelastic model for the myocardium}\label{sect:proposed_vemodel}

In this paper, we propose a viscoelastic model capable of capturing the behavior of myocardial tissue. 
Building from sections~\ref{sect:he} and~\ref{sect:ve}, we propose the model form,
\begin{equation}
    \bs{S}(t) = \bs{S}_{e} (t) + \bs{S}^\star(t) + \bs{S}_{p} (t).
    \label{eq_frac_zener_model2}
\end{equation}
In this case, the elastic form $ \bs{S}_e$ denotes the response of the underlying base structure of the tissue and is characterized by a simple neo-Hookean model with a single stiffness parameter, a, \emph{e.g.}
\begin{equation}
    \bs{S}_e = \frac{\text{a}}{J^{2/3}} \left ( \bs{I}  - \frac{\I_{\C}}{3} \C^{-1} \right ). 
\end{equation}
The complex nonlinear and viscoelastic response of the tissue is incorporated through $ \bs{S}^\star $, which is governed by the fractional differential equation
\begin{equation}
    \bs{S}^\star(t)  
    + 
    \delta D_t^{\alpha} \bs{S}^\star
    = 
    D^\alpha_t \bs{S}_{v}.
    \label{eq_frac_zener_model1}
\end{equation}
In this formulation, the underlying material response is dictated by the choice of $ \bs{S}_v $, with the viscoelastic effects controlled by the parameters $ \alpha $ and $ \delta $.
Mimicking the structural forms of both \HO and Costa models, we choose $ \bs{S}_v$ to be,
\begin{equation}
    \bs{S}_v 
    =
    \mathcal{W}_1
        \sum_{\text{kl} \in S} 
        \text{b}_\text{kl} \Ev{k} \otimes \Ev{l}
    +
    \mathcal{W}_2 
        \sum_{\text{kl} \in S^\perp} 
        \text{b}_\text{kl} \I_\text{kl}  \text{sym} (\Ev{k} \otimes \Ev{l}),
    \label{eq:Sv}
\end{equation}
where $ S = \{ \text{ff}, \text{ss}, \text{nn} \} $ and $ S^\perp = \{ \text{fs}, \text{fn}, \text{sn} \}$ and
\begin{align*}
    \mathcal{W}_1(\C) 
    & = 
    \exp \{ \text{b}_1 (\I_{\C} - 3) \},
    \nonumber \\
    \mathcal{W}_2(\C) 
    & = 
    \exp \{ \text{b}_2 (\I_\text{fs}^2 + \I_\text{fn}^2 + \I_\text{sn}^2 ) \}.
    \nonumber 
\end{align*}
Like the \HO model, the first term contains an isotropic scaling, $ \mathcal{W}_1$; however, this term scales weighted structurally anisotropic terms.
As a hyperelastic contribution, this term would lead to a diagonal matrix at the reference state, suggesting the reference state is not a stable state.
However, as part of the fractional viscoelastic model (differentiated by the Caputo derivative), this is no longer a factor.
Similar to the Costa model, the invariant terms are grouped, with the first term covering the sum of the diagonal invariants (note, $ \I_{\C} = \I_\text{ff} + \I_\text{ss} + \I_\text{nn} $) and the second incorporating the shear invariants.
Similar to the first term, $ \mathcal{W}_2 $ scales weighted shear terms.
Unlike the Costa model, weighting is done through scalings $ \text{b}_\text{kl} $ which are not reflected in the exponents of $ \mathcal{W}_1  $ or $ \mathcal{W}_2 $.
While incorporation of scalings $ \text{b}_\text{kl} $ into the exponential scalings of $ \bs{S}_v $ would make it straightforward to write as a strain-energy function, given the model is viscoelastic, this constraint is lifted.
Moreover, grouping terms with two exponents, $ \text{b}_1 $ and $ \text{b}_2 $, significantly simplifies the parameterization process.
Last, the model considered in this paper utilizes an incompressible formulation, where $ \bs{S}_p = pJ\C^{-1} $. However, other formalisms are possible to incorporate compressibility (though care must be taken when considering these forms~\cite{Nolan2014}).

Important properties for constitutive models are objectivity and material frame indifference~\cite{wineman2009nonlinear}. 
While these properties are well-known for $ \bs{S}_e$ and $ \bs{S}_p$, these properties must hold for $ \bs{S}^\star$.
Examining Eq.~\eqref{eq_frac_zener_model1}, it's  sufficient to require that these properties hold for $ \bs{S}_v $. 
Specifically, that $ \bs{S}_v $ is invariant to rigid (potentially dynamic) rotations of the physical frame and that, given an orthonormal rotation $ \bs{H} $ of the material frame,
\begin{equation}
    \bs{H} \bs{S}_v ( \C, \Ev{m} \otimes \Ev{n}, \ldots ) \bs{H}^T
    =
    \bs{S}_v ( \bs{H} \C \bs{H}^T , \bs{H} \Ev{m} \otimes \Ev{n} \bs{H}^T,  \ldots).
    \label{eq_matframe}
\end{equation}
Objectivity holds as $ \bs{S}_V $ is a strict function of the invariants of $ \C $.
Further, material frame rotations leave all invariants unchanged and, due to rotations of the microstructural directions, $ \bs{S}_v $ satisfies Eq.~\eqref{eq_matframe}.

The proposed viscoelastic model in Eq.~\eqref{eq_frac_zener_model2} requires $ 11 $ parameters in total.
Seven constants are used to scale the isotropic and anisotropic model contributions linearly,  $\bs \theta^l = \{ \text{a}_\text{e} $ $\text{b}_{\text{ff}}$,  $\text{b}_{\text{ss}}$, $\text{b}_{\text{nn}}$,  $\text{b}_{\text{fs}}$, $\text{b}_{\text{fn}}$, $\text{b}_{\text{sn}}  \} $. 
The remaining four parameters -- including the fractional order, the scaling on the stress in Eq.~\eqref{eq_frac_zener_model1}, and the exponential scaling parameters -- have a nonlinear dependence and contribute to the nonlinear and viscoelastic response of the material $\bs \theta^* = \{ \alpha, \: \delta, \: \text{b}_\text{1}, \: \text{b}_\text{2} \}$.

\begin{figure}[t!]
\center
\subfloat[triaxial shear deformation modes]{\includegraphics[width=0.48\textwidth]{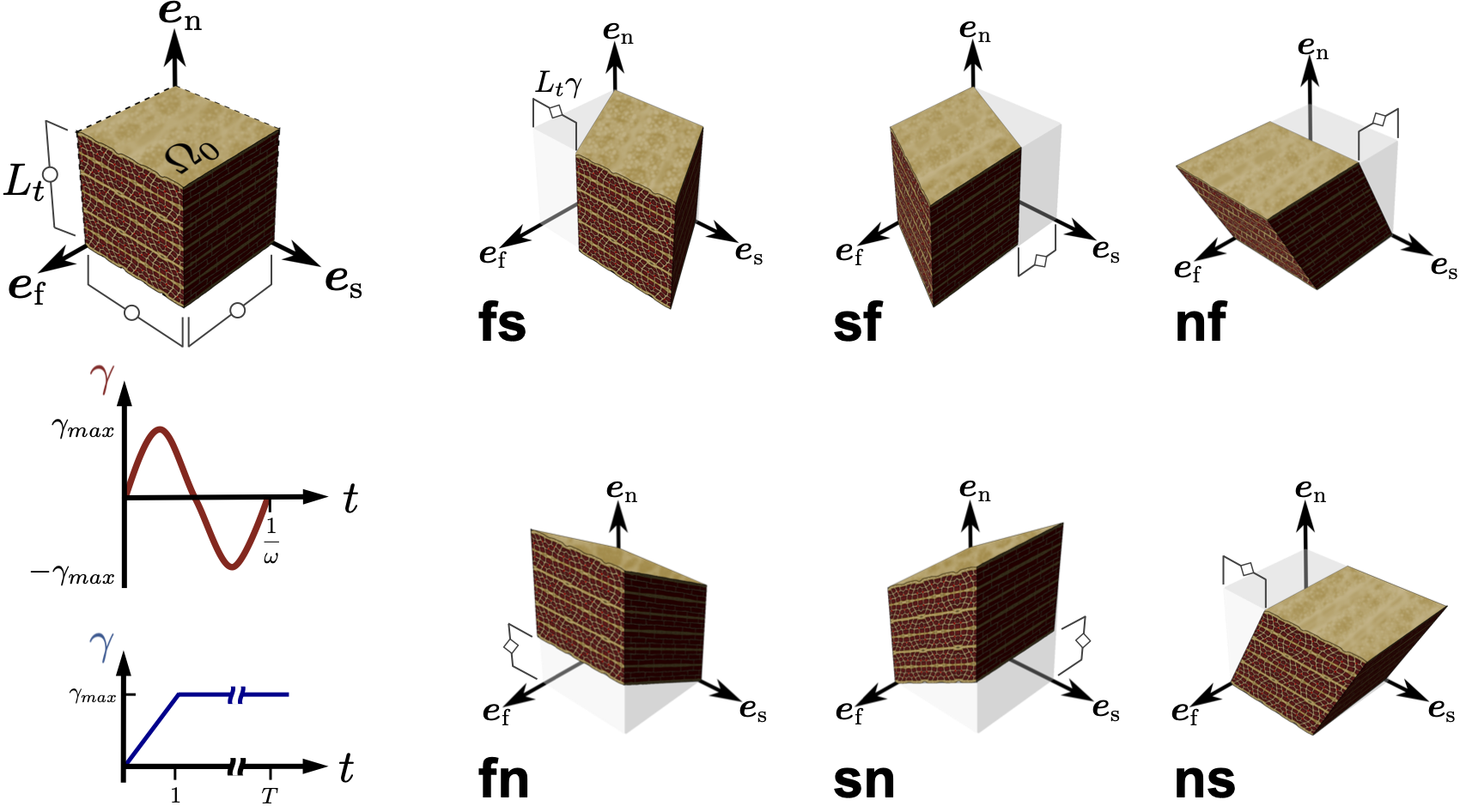}\label{F:exp_tri}} \\ 
\subfloat[biaxial stretch deformation mode]{\includegraphics[width=0.48\textwidth]{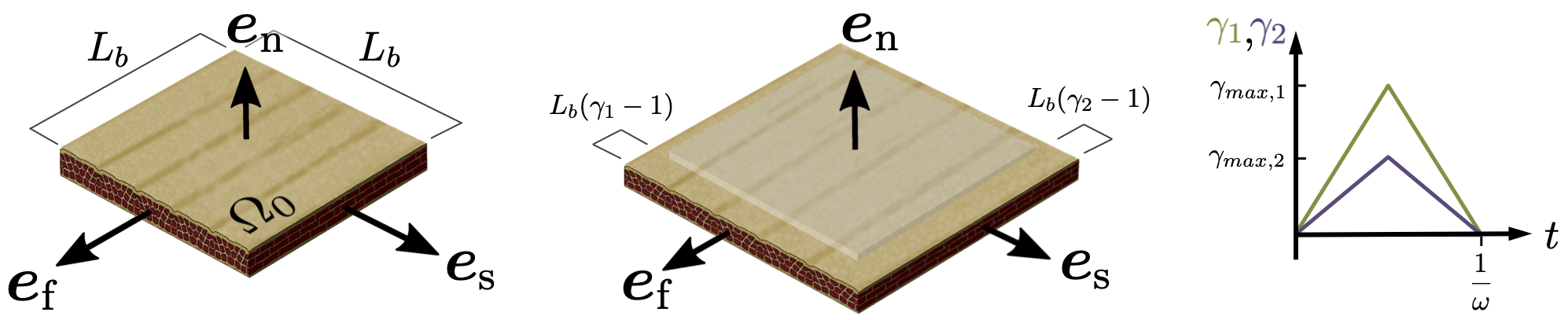}\label{F:exp_biaxial}} \\
\subfloat[transient deformation protocols]{\includegraphics[width=0.48\textwidth]{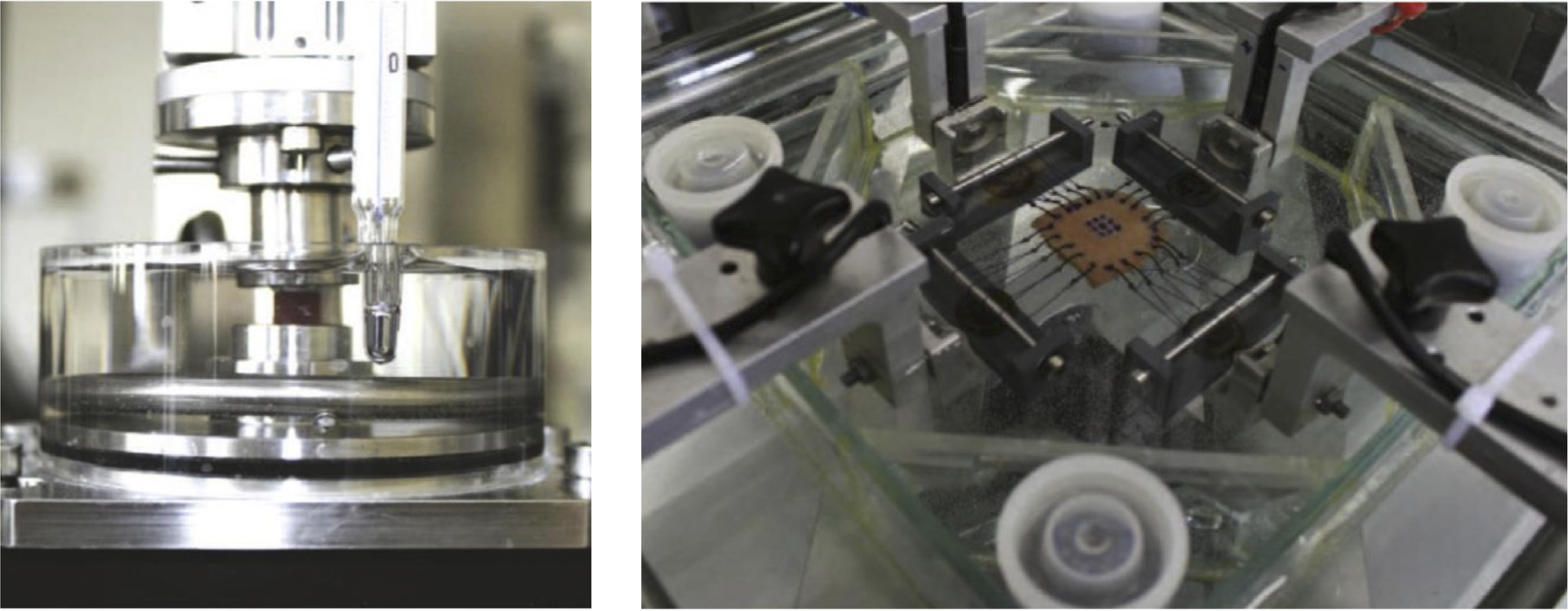}\label{F:exp_graph}}
\caption{Illustration of the experimental tests performed on human myocardial samples in~\cite{sommer2015biomechanical}.
(a) Reference tissue block with shear deformations applied in different directions relative to the underlying tissue microstructure.
(b) Reference biaxial sample with stretch applied in fiber and sheet directions.
(c) Experimental test rigs for triaxial and biaxial experiments.
} \label{F:exp_general}
\end{figure}

\subsection{Experimental testing of human myocardium} \label{sect:data}

This modeling study relies on rheological tests performed on human myocardial samples by Sommer \emph{et al.}~\cite{sommer2015biomechanical,sommer2015quantification}.
While complete details on the experimental data acquisition can be found in~\cite{sommer2015biomechanical,sommer2015quantification}, we reiterate the basics of data acquisition here (see figure~\ref{F:exp_general}).
Briefly, human heart muscle samples were collected during transplant surgery, infused with 200mL cardioplegic solution (CPS, Celsior by Genzyme Corporation), inserted into a path of 1000 mL of CPS and cooled to $ 4^\circ $C.
From the collected hearts, samples were cut into $ 25 \times 25 \text{ mm}^2$ thin sheets for biaxial testing and $ (4 \text{ mm})^3$ cubed samples for shear testing.
Passive tests of the cardiac muscle were performed at $37^\circ$C, in a bath of CPS with 20 mM 2,3-butanedione monoxime (BDM) .

Biaxial extension tests were performed on thin square specimens whose sides were aligned with the mean-fiber ($\Ev{f}$) and sheet directions ($\Ev{s}$) (figure~\ref{F:exp_biaxial}). 
Equibiaxial stretches ranging between 5--20\% were performed, as well as non-equibiaxial stretches of varying ratios between the mean-fiber and sheet directions, $(\Ev{f}:\Ev{s}) \in \{$(1:1), (1:0.75), (1:0.5), (0.75:1), (0.5:1)$\}$. Four preconditioning cycles were performed before acquiring the data during the fifth cycle, at a loading speed of 3 mm/min. 

Triaxial shear tests were performed on cubes extracted from adjacent locations of the biaxial samples. 
The cube sides were cut in order to align with the tissue microstructure (figure~\ref{F:exp_tri}). 
Shear deformations between $10$ and $50\%$ were performed, with a $10\%$ increment and loading speed of 1 mm/min. 
Shear relaxation tests were also performed at 50\% at a ramp speed of 100 mm/min and a duration of 5 min. 
As a cube can be used to test two orthogonal directions, a total of three specimens were needed to test all six modes of simple shear. 
For triaxial testing, two deformation cycles were used for preconditioning, with the data being acquired during the third cycle.

In total, $ 14 $ tests were used to fit models: 6 shear relaxation tests at $ 50\%$ shear, 6 cyclic shear tests at $ 50\%$ shear maximum amplitude, and fiber/cross fiber equibiaxial stretch at $10 \%$. 
To verify the models, the identified parameters were used to predict the behavior of myocardium in non-equibiaxial stretch tests (fiber to sheet stretch ratios of $ \{$(1:1), (1:0.75), (0.75:1), (1:0.5), (0.5:1)$\}$) and in equibiaxial stretch at three different frequencies: 0.01, 0.033 and 0.1 Hz. 
Additionally, the model prediction at lower shear levels (10-40\%) is compared against data acquired using two different protocols -- increasing or decreasing shear levels.

\subsection{Parameter identification and analysis}\label{sec:paramid}

The  proposed model parameters were fit to human triaxial and biaxial experimental data collected in~\cite{sommer2015biomechanical}, assuming idealized shear and biaxial kinematics, \emph{i.e.}
\begin{equation*}
    \F = \gamma \Ev{k} \otimes \Ev{l} + \bs{I},
    \; 
    \text{ and }
    \;
    \F 
    = 
    \gamma_1 \Ev{f} \otimes \Ev{f} 
    + 
    \gamma_2 \Ev{s} \otimes \Ev{s}
    + 
    \frac{1}{\gamma_1 \gamma_2} \Ev{n} \otimes \Ev{n},
\end{equation*}
where $ \Ev{k}, \Ev{l} $ denote shear along different microstructural directions, and $ \gamma, \gamma_1, \gamma_2 $ denote the time dependent amount of stretch or shear applied for each test (see figure~\ref{F:exp_general}).
Here we refer to the various data tests as groups - relaxation, cyclic shear and biaxial stretch, which will be indicated by superscripts $i \in \{r,c,b\}$, respectively. 
The relaxation and cyclic shear groups comprise six tests each, indicated by subscript $ kl $ corresponding to the shear directions $ \text{kl} \in M^s=\{$fs, fn, sf, sn, nf, ns$\}$, while the biaxial stretch group comprises 2 tests indicated by subscript $ \text{kl} \in M^b=\{$ff, ss$\}$. 
Hence, a total of 14 tests were used for model fitting.
For each of those, let $\sigma^i_\text{kl}$ denote the respective component of the Cauchy stress, with $\bar{\sigma}_\text{kl}$ marking the values measured in the experiments, and $\sigma$ the values computed using the model.
The bold versions, $ \bs{\sigma}^i_\text{kl} $ and $ \bs{\bar{\sigma}}^i_\text{kl} $, denote the stress over time for the $ i^{th} $ test set and $ \text{kl}^{th} $ stress component.
Note, all models were preconditioned following the data protocol, and compared at their final cycles (where applicable).

\subsubsection{Minimisation problem}\label{sec:minimisation}

In this paper, we consider the fit of three models: the \HO model, Costa model, and proposed viscoelastic model from section~\ref{sect:proposed_vemodel}.
All models considered were fit to data by minimizing the objective function across all 14 tests simultaneously (Eq.~\eqref{eq_param_min}) to determine the model's set of $N$-parameters (denoted as $ \bs \theta $)  
\begin{equation}
    \bs{\theta}
    =
    \argmin_{ y \in \mathbb{R}^N_+ } \left \{ \; \min_{\bs{\beta} \in \mathbb{R}^3}
    \chi (\bs{y}, \bs{\beta}) \; \right \},
    \label{eq_param_min}
\end{equation}
where the objective function, $ \chi $, is given as
\begin{equation}
    \chi (\bs{y}, \bs{\beta})
    =
    \dfrac{ \left ( \sum_{i,kl} \left \| \frac{1}{R_{kl}^i} \left(
        \frac{1}{\beta^i} \bs{\sigma}_{kl}^i(\bs{y})  
        -  
        \bs{\bar{\sigma}}_{kl}^i
        \right) \right \|_2^2 \right )^{1/2 }}
    {\left ( \sum\limits_{i,kl} \left \| 
        \frac{1}{R_{kl}^i} \bs{\bar{\sigma}}_{kl}^i
        \right \|_2^2 \right )  ^{1/2} },
    \label{eq_objfunc}
\end{equation}
and $ R_{kl}^i = \| \bs{\bar{\sigma}}_{kl}^i \|_2 $ is introduced to give tests equivalent weights (irrespective of the magnitude or number of time points). 
Note that $ \chi $ gives the relative error across all tests, with $ \chi = 1$ denoting $ 100 \%$ error.
In this minimization, a parameter set is found to best match all datasets.

In this study, the data is representative of the myocardium behavior in relaxation, cyclic shear and biaxial stretch, but it is not guaranteed to come from the same sample (e.g., for the same shear mode in relaxation and cyclic oscillations), or even from the same heart. 
Therefore, innate variability is introduced that may be reflective of animal to animal variability.
Here we assume that the relative magnitudes test to test may vary as a result of different samples, but the shapes should be maintained.
This is achieved through the introduced relative scalings $\bs{\beta}$, which are selected to minimize the difference between model and data across all time points. 
In order to meaningfully preserve the anisotropy, $ \bs{\beta} = ( \beta^r, \beta^c, \beta^b )$ was applied across all $ kl $ directions for given tests.
The selection of $ \bs{\beta} $ was done iteratively with an initial guess $ \bs{\beta} = \bs{1} $, solving the minimization for $ \bs{y} $, minimizing for $ \bs{\beta} $ and repeating until convergence.

Errors were also quantified on a test by test basis to understand areas of model strength and weakness using the related test-specific objective function,
\begin{equation}
    \chi_{kl}^i (\bs{\theta})
    =
    \dfrac{ \left \| \frac{1}{R_{kl}^i} \left(
        \frac{1}{\beta^i} \bs{\sigma}_{kl}^i(\bs{\theta})  
        -  
        \bs{\bar{\sigma}}_{kl}^i
        \right) \right \|_2 }
    {\left \| 
        \frac{1}{R_{kl}^i} \bs{\bar{\sigma}}_{kl}^i
        \right \|_2 },
    \label{eq_objfunc_test}
\end{equation}
where the $ \beta^i $ and $ \bs{\theta} $ are determined from the minimization in Eq.~\eqref{eq_param_min}.
Minimization for the Costa and \HO models were done using the nonlinear \texttt{lsqnonlin} minimization routine in MATLAB. 
Multiple initial guesses were selected for the model parameters, with the reported results selected using the best fit.

To examine the behavior and uniqueness of the parameter space of the proposed viscoelastic model, a parameter sweep was performed.
To generate model predictions, each experiment had to be simulated based on the given kinematics to solve the fractional differential equation forward in time (see~\cite{zhang2020efficient,zhang2021compare} for details).
In this case, we exploited the linearity of some model parameters $ \bs{\theta}^l$, which could be found through a simple least squares matrix solve for a given the set of nonlinear parameters $ \bs{\theta}^*$; reducing the parameter space from $ \mathbb{R}^{11}_+ $ to $ \mathbb{R}^4_+ $.
Uniqueness of the linear parameters was then determined by the solvability of the least squares system.
Because of the simple kinematics and known stresses during the relaxation tests, it was also possible to reduce the parameter space to $ \mathbb{R}^3_+$ by determining the functional relationship between $ \delta $ and $ \alpha $ during the relaxation tests. 
Sweeps were performed over $ \text{b}_1 \in \{ 0,1,2 \ldots 20 \}$, $ \text{b}_2 \in \{ 0,0.5,1 \ldots 10 \} $, and $ \alpha \in \{ 0.2, 0.22, 0.24 \ldots 0.4 \} $ resulting in $4,851$ minimizations and $67,914$ test simulations.
Note, a coarser sweep over higher parameter values showed continued growth in $ \chi $.
This sweep indicated that the minimum is achieved at $\alpha=0.24$, $\text{b}_\text{1}=9$ and $\text{b}_\text{2}=2$.
A refined sweep was then conducted around these values, as follows: $\alpha \in \{ 0.2, 0.21, 0.22 \ldots 0.28 \} $, $\text{b}_\text{1} \in \{ 7, 7.5, 8 \ldots 11 \} $ and $\text{b}_\text{2} \in \{1, 1.25, 1.5 \ldots 3.5 \}$.

\subsubsection{Model sensitivity analysis}\label{sec:param_sensi}
Sensitivity was assessed from two perspectives: the effect of noise in the data on the best-fit material parameters, and the effect of parameter perturbations on the fitting error.
In the first instance, 100 noisy datasets were produced via adding unbiased uniformly distributed noise vector over time $ \bs{\eta} $, to the measured stress values: $\bs{\tilde{\sigma}}^i_{kl} = \bs{\bar{\sigma}}^i_{kl} + \bs{\eta} $. 
The noise level was set to $10\%$ of the peak stress across a test for a specific mode, \emph{e.g.}, $ \eta \in 0.1  [-|\bs{\bar{\sigma}}^i_{kl}|_\infty,|\bs{\bar{\sigma}}^i_{kl}|_\infty] $, $ E(\bs{\eta}) = 0 $.
For each noisy dataset $\bs{\tilde{\sigma}}^i_{kl} $, a model fit with a different set of optimal parameters $\bs \theta $ were obtained, following the same parameter fit procedure.
Next, the optimal parameters $\bs \theta_{\min}$ were perturbed by $\pm 10\%$ of the original value, by turn. 
The error function $ \chi $ was computed for each local perturbation, according to Eq.~\eqref{eq_objfunc}.

\begin{figure*}[ht!]
\centering
\includegraphics[width=1\linewidth]{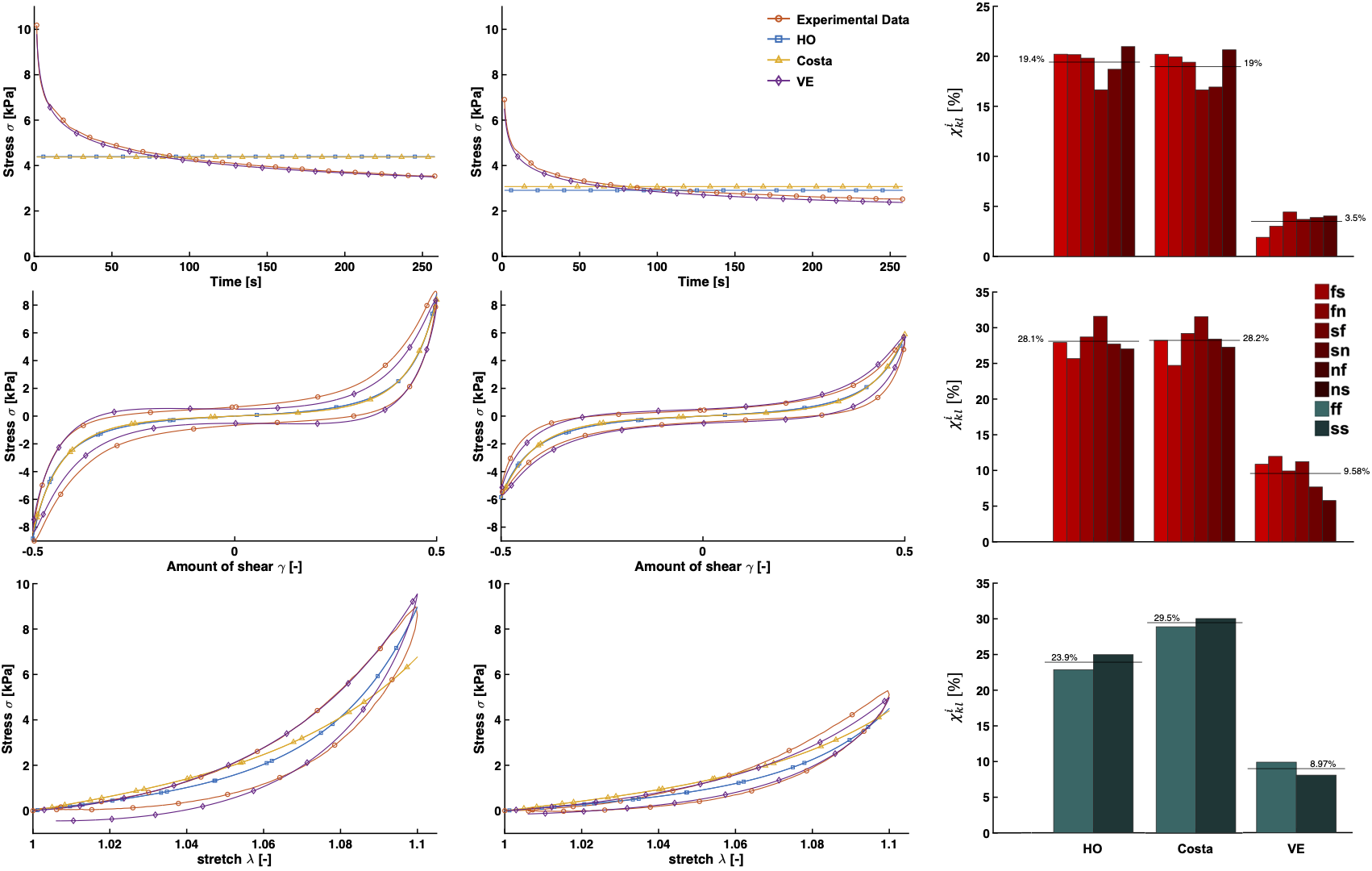}
\caption{
Constitutive model fit for \HOns, Costa and the proposed viscoelastic (VE) models, with overall error ($ \chi $) of $ 24.17  $, $ 24.95  $, and $ 7.65 \% $, respectively.
Optimal parameters for \HO were $ \{ \text{a}_\text{f}$, $\text{a}_\text{s}$,  $\text{a}_\text{fs}$,  $\text{a}$, $\text{b}_\text{f}$, $\text{b}_\text{s}$,  $\text{b}_\text{fs}$, $\text{b} \}  = \{ 1.56, 0.70, 0.46, 0.61, 35.31, 33.24, 5.09, 7.50 \} $, with \emph{a} parameters given in units of \texttt{kPa} and $ \{ \beta^r, \beta^c, \beta^b \} = \{1.41, 0.74, 0.53 \} $.
Optimal parameters for Costa were $ \{ $C, b$_\text{ff}$, b$_\text{ss}$, b$_\text{nn}$, b$_\text{fs}$, b$_\text{fn}$, b$_\text{sn} \} = \{0.13, 33.27, 20.83, 2.63, 12.92, 11.99, 11.46 \} $, with C in units of \texttt{kPa} and $ \{ \beta^r, \beta^c, \beta^b \} = \{1.37, 0.75, 0.14 \} $.
Optimal parameters for VE were $ \{ \alpha $, $\delta$, b$_1$, b$_2 \} = \{0.23, 0.47, 9.5, 2.25 \} $, $ \{ \text{a}_\text{e}$, b$_\text{ff}$, b$_\text{ss}$, b$_\text{nn}$, b$_\text{fs}$, b$_\text{fn}$, b$_\text{sn} \} = \{0.54, 3.34, 1.80, 0.77, 6.62, 3.64, 2.93 \} $
with a$_\text{e}$ and $\text{b}_\text{mn}$ in units of \texttt{kPa} and $ \{ \beta^r, \beta^c, \beta^b \} = \{1.42, 0.82, 0.30 \} $.
Model fits shown for shear relaxation (top left/center) and cyclic shear (middle left/center) in fs and sf directions and biaxial stretch (bottom left/center) in ff and ss directions.
Test-specific errors, $ \chi^{i}_\text{mn}$, are shown for the relaxation (top right), cyclic shear (middle right) and biaxial stretch (bottom right) tests for all models.}\label{F:Model_fit}
\end{figure*}

\section{Results}~\label{sect:results}

\begin{figure}[ht!]
\center
{\includegraphics[width=\linewidth]{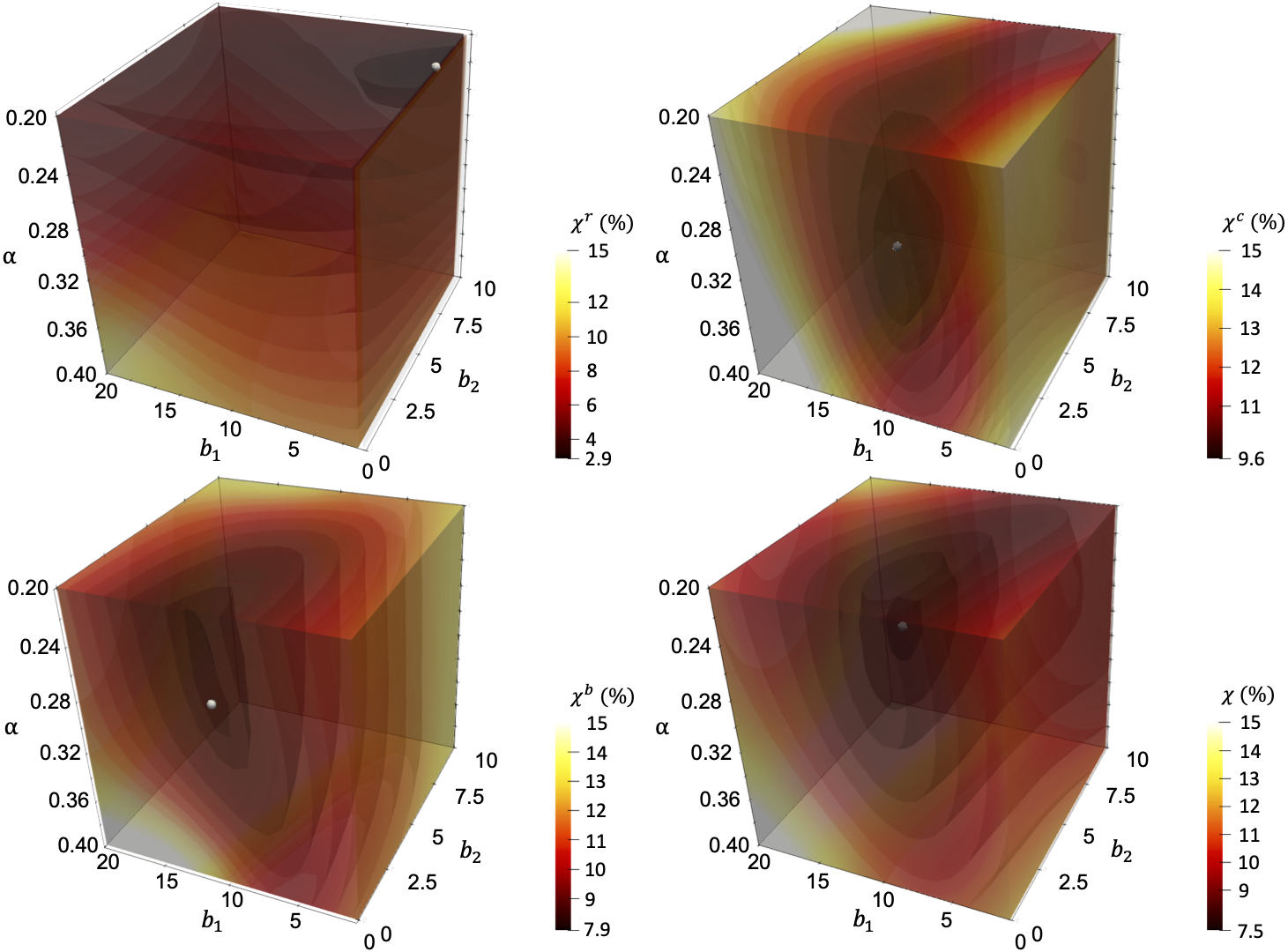}} \quad
\caption{Values of the objective function $ \chi $ across fractional order values $\alpha \in[0, \: 1]$ and exponential powers $b_1, \: b_2 \in [0, \: 20]$. The heat maps indicate how the error changes across the 3D space. Isosurfaces are aiding in the visualisation of the error, with the white spheres indicating the location where the minimum is achieved.
(Top Left) Error $\chi^r$ across the relaxation tests only. The minimum (2.86\%) is achieved at $\alpha=0.2$, $b_1=1$ and $b_2=7$. 
(Top Right) Error $\chi^c$ across the cyclic shear tests only. The minimum (9.65\%) is achieved at $\alpha=0.28$, $b_1=8.5$ and $b_2=1$.
(Bottom Left) Error $\chi^b$ across the biaxial stretch tests only. The minimum (7.90\%) is achieved at $\alpha=0.26$, $b_1=10$ and $b_2=0$.
(Bottom Right) Error $ \chi $ across all tests. The minimum (7.65\%) is achieved at $\alpha=0.23$, $b_1=9.5$ and $b_2=2.25$.}\label{F:err_alpha}
\end{figure}

Figure~\ref{F:Model_fit} shows the best fit for \HOns, Costa and proposed viscoelastic (VE) models to the data. 
Example curves for the relaxation, cyclic shear and biaxial tests are shown along with bar plots showing test-specific errors for each of the 14 datasets, computed according to Eq.~\eqref{eq_objfunc_test}. 
The overall errors, computed using Eq.~\eqref{eq_objfunc}, were $ 24.17 $, $ 24.95 $ and $ 7.65 \% $ for \HOns, Costa and VE models, respectively.
In addition, figure~\ref{F:err_alpha} explores the VE model error across the $\alpha$ (0.2 to 0.4), $\text{b}_1$ (0 to 20) and $\text{b}_2$ (0 to 10) parameter space. 
The error across all datasets, computed according to Eq.~\eqref{eq_objfunc}, is shown in the bottom right panel. 
The error computed separately for the relaxation, cyclic shear and biaxial tests is shown in the top left, top right and bottom left panels, respectively. The isosurfaces help visualizing the error behavior in the 3D space.
In each of the four panels, the minimum error is indicated by the white sphere with the overall minimum uniquely identified at $\alpha=0.23$, $\text{b}_1=9.5$ and $\text{b}_2=2.25$ over all datasets. The errors over the individual tests were computed using Eq.~\eqref{eq_objfunc_test}.

\begin{figure}[h!]
{\includegraphics[width=\linewidth]{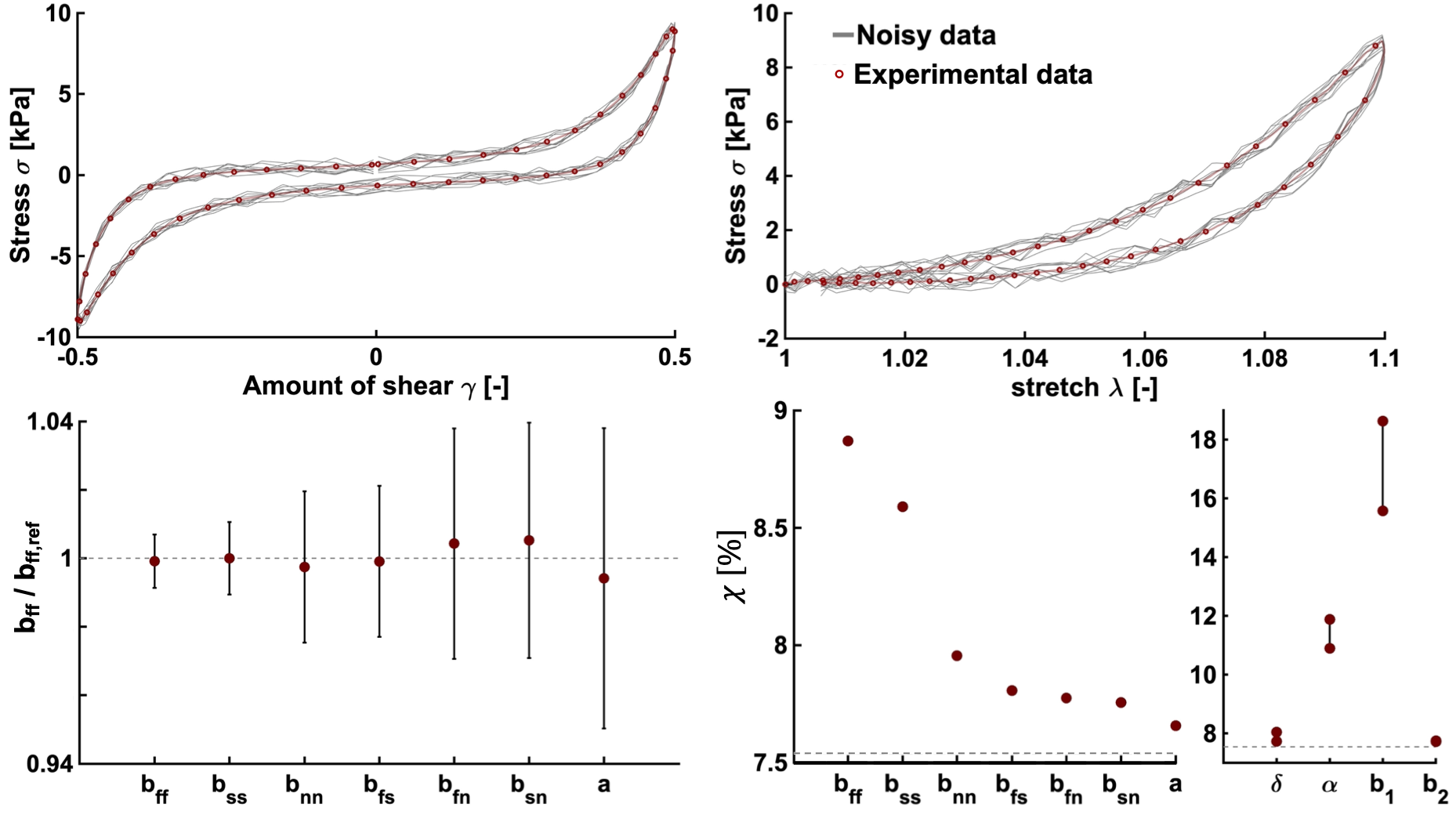}} 
\caption{Sample datasets for the fs and ff modes of deformation with added $10\%$ unbiased noise compared to the original experimental data. (Top Left) Noisy data in fs cyclic shear and (Top Right) ff biaxial stretch. (Bottom Left) Relative variation of model parameters obtained by fitting 100 noisy datasets. For each parameter the mean (shown as the red dot) and the range of one standard deviation (solid black line) can be compared relative to the original value (dotted line). 
(Bottom Right) Variation of the objective function $ \chi $ due to $10\%$ variation in parameters.
} \label{F:Noisy}
\end{figure}

Figure~\ref{F:Noisy} illustrates synthetic noisy datasets alongside the original data, showing the resultant noisy data for cyclic shear (fs, top left) and biaxial (top right) stretch (ff).
Equivalent datasets were generated for all test cases, and replicated 100 times to examine the sensitivity of the fit to noise.
For each of the 100 noisy datasets, a new set of VE linear parameters $\bs \theta^l$ were determined.
The parameters' mean (red marker) and standard deviation (error bars) are shown relative to the original parameters, indicated by the baseline 1 (bottom right).
To further understand model sensitivity to parameters, variation of the objective function was found when changing parameters by $\pm$10\% (bottom left) for both linear and nonlinear parameter sets.
The original parameter values can be seen in figure~\ref{F:Model_fit}. The dotted line indicates the minimum model fit error of $7.65\%$. 

\begin{figure}[h!]
\center
\includegraphics[width=\linewidth]{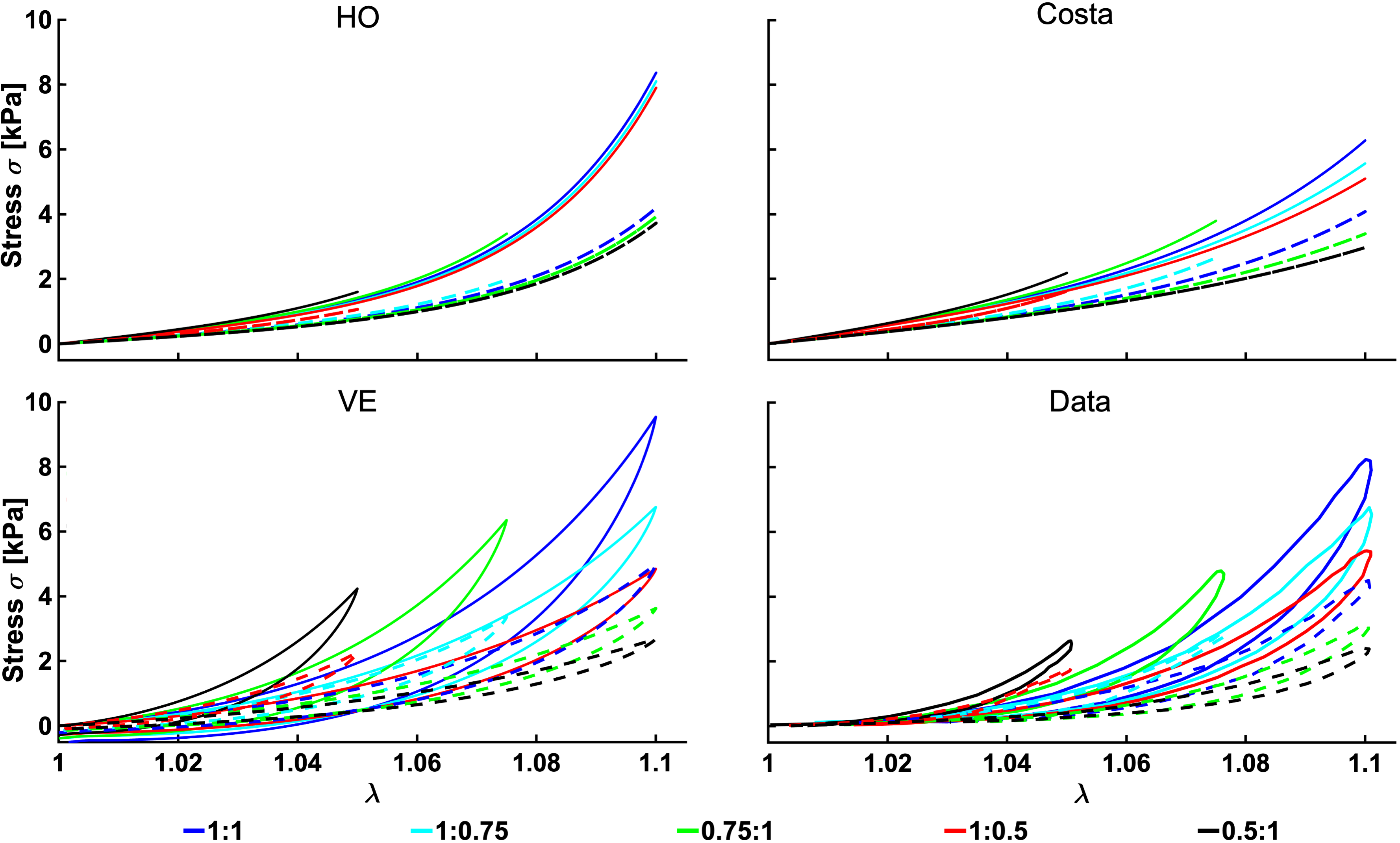}
\caption{Prediction of the models in biaxial stretching, fiber to sheet ratio ff:ss of 1:1 (blue), 1:0.75 (cyan), 0.75:1 (green), 1:0.5 (red) and 0.5:1 (black). Solid curves show the behavior in the fiber direction, and dashed curves in the sheet direction. (Top left) \HO model prediction with $\beta^b=0.65$; (Top right) Costa model prediction with $\beta^b=0.14$; (Bottom left) VE model prediction with $\beta^b=0.30$; (Bottom right) Data from the original study~\cite{sommer2015biomechanical}. Note all model predictions used parameters from figure~\ref{F:Model_fit}.} \label{F:prediction_biax}
\end{figure}

Figure~\ref{F:prediction_biax} shows the prediction of the models under biaxial stretches of various ff:ss ratios (1:1 -- blue, 1:0.75 -- cyan, 0.75:1 -- green, 1:0.5 -- red and 0.5:1 -- black). 
The models are shown in the top left (\HOns), top right (Costa) and bottom left (VE) panels, while the original data, from~\cite{sommer2015biomechanical}, are presented in the bottom right panel. 
The solid curves indicate the fiber response, and the dashed curves show the response in sheet direction. 
Note that here none of the models was fit to data (except for ff:ss ratio of 1:1 which was part of the original fitting dataset), and the curves represent predictions obtained by using the parameters shown in figure~\ref{F:Model_fit}.
 
\begin{figure}[h!]
\center
\includegraphics[width=\linewidth]{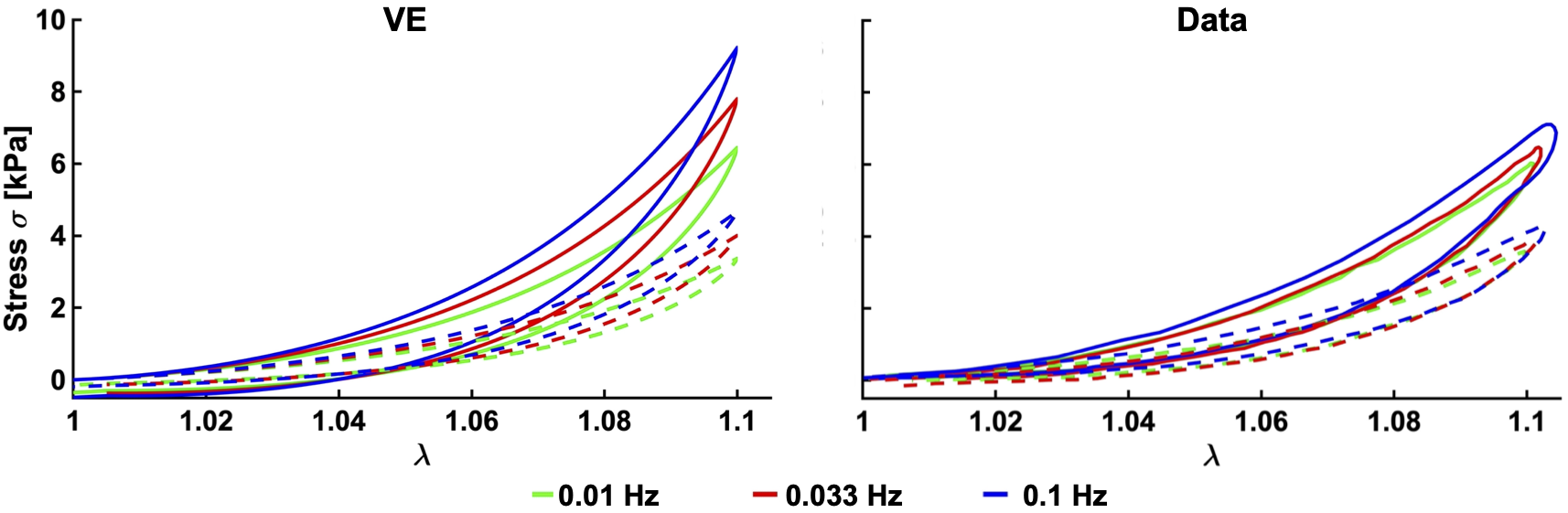}
\caption{The behavior of the proposed VE (left) against data from the original study~\cite{sommer2015biomechanical} (right), in biaxial stretch tests with variable loading frequency (0.01 Hz -- green, 0.033 Hz -- red, 0.1 Hz - blue). Solid curves show the behavior in the fiber direction, and dashed curves in the sheet direction. The model employs the parameters shown in figure~\ref{F:Model_fit} and $\beta^b=0.45$.} \label{F:VE_prediction_freq}
\end{figure}

Figure~\ref{F:VE_prediction_freq} shows the prediction of the viscoelastic  model in ff:ss biaxial stretch (left), at three frequencies: 0.01, 0.033 and 0.1 Hz. The corresponding data, from the original study~\cite{sommer2015biomechanical}, is shown in the right panel. 
At 0.01 Hz, the testing conditions are identical to the biaxial test used for the model fitting. 
However, the data used for the fitting and the data shown in figure~\ref{F:VE_prediction_freq} come from different samples, with the data in the frequency test reaching fiber and sheet peak amplitude smaller by a factor  $\sim$1.48.
 
\begin{figure}[h!]
\center
\includegraphics[width=\linewidth]{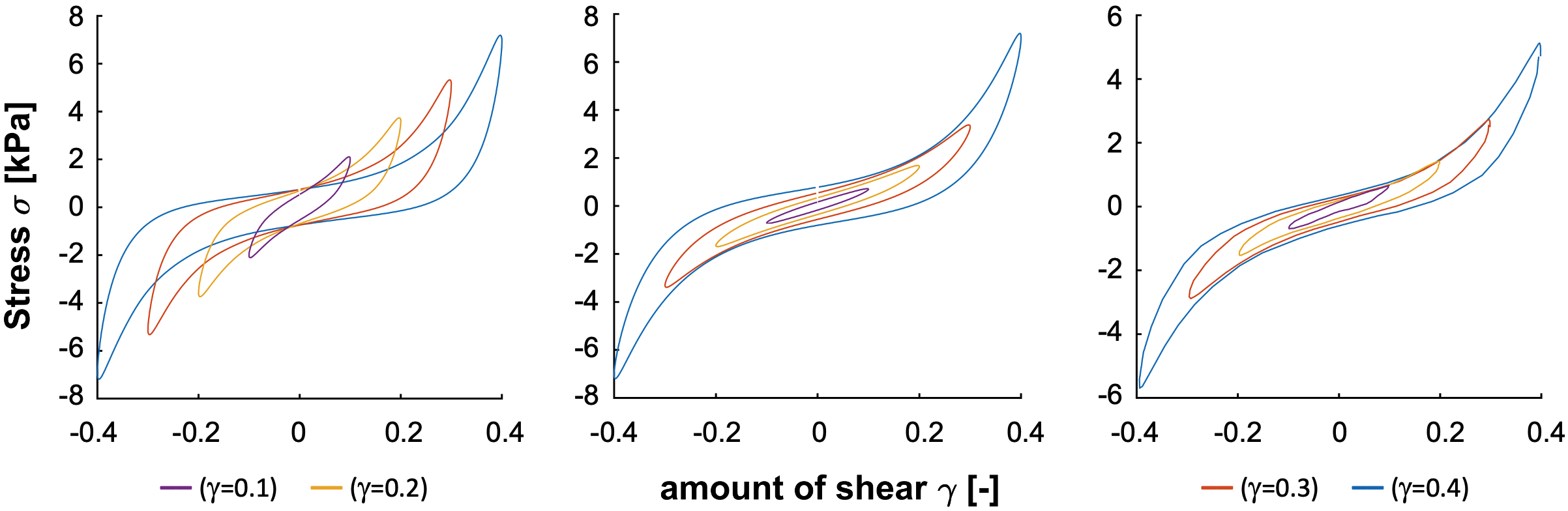}
\caption{Qualitative comparison between data and model in cyclic shear. 
(Left) shear at multiple levels from human myocardium~\cite{sommer2015biomechanical}, (center) model predictions of shear response, and (right) cyclic shear in porcine myocardium.
Experiments used the same rig and protocols, with the only exception that (left) increased shear levels (preconditioning each level), while (right) started from higher shear, decreasing shear levels.} \label{F:shear_prediction}
\end{figure}

Figure~\ref{F:shear_prediction} shows the VE model prediction to shear compared with data from the original study~\cite{sommer2015biomechanical} on human myocardium as well as porcine myocardium.
The shear tests conducted for human myocardium start from low ($\gamma=0.1$) to high shear ($\gamma=0.4$), while the data from porcine myocardium start from high shear ($\gamma=0.4$) to low ($\gamma=0.1$).
In both cases, 4-6 cycles are performed at a given shear level with the final cycle shown. 
Predictions from the VE model are shown (center) using the parameters presented in figure~\ref{F:Model_fit} and scaled by $\beta^c=0.38$, which was determined by matching the model to original data (top row) peak amplitude in the cyclic shear test at $\gamma=0.4$ in the human myocardial data.
 
\section{Discussion}\label{sect:discussion}

\subsection{Analysis of the proposed viscoelastic model}

The model proposed in this work introduces viscoelasticity through a fractional approach. 
This choice reflects a hierarchy of relaxation times mirroring the hierarchical structure of myocardial tissue and the different spatiotemporal scales that lead to viscous dissipation.
The proposed viscoelastic model presented here relies on an underlying form that combines aspects of the \HO and Costa hyperelastic models. 
As in the \HO model, the model terms used here are based on invariants that reflect the microstructural composition of the passive myocardium. 
However, the terms are not completely separated, allowing for coupling between invariants, as seen in Eq.~\eqref{eq:Sv}. 
Coupling of the microstructural directions is characteristic of the Costa model, where all stretches are inherently coupled.
Overall, the VE model captures the characteristics of the data, as seen in representative examples of the three deformation protocols in figure~\ref{F:Model_fit}. 
In relaxation, the model can capture the initial peak and the subsequent decay. 
In cyclic shear and biaxial stretch, the hysteresis and nonlinearity are well matched.

The viscoelastic model entails 11 parameters uniquely determined through the fitting process. 
Four of these have a nonlinear effect on the model - the fractional order $\alpha$, exponential powers $\text{b}_1$ and $\text{b}_2$ and the viscoelastic PK2 scaling $\delta$, while the remaining 7 parameters act as linear scalings of terms. 
Parameter $\delta$ is determined from the relaxation tests, at each $\alpha$ value, as explained in section~\ref{sec:paramid}. 
For the other 3 nonlinear parameters, a sweep is carried, and at each point in the 3D space a set of 7 unique linear parameters can be identified. 
Figure~\ref{F:err_alpha} shows the error behavior across the 3D space of the sweep. Overall, it can be observed that the region of minima for the cyclic shear and biaxial tests (top right, bottom left) is close to that of the entire dataset error (bottom right), with $\alpha$ between 0.23 and 0.28, intermediate b$_1$ and small b$_2$. 
For the relaxation tests, the $\alpha$ value is similar, yet b${_1}$ is small and b${_2}$ is large. 
However, of all groups, the best parameter region in the relaxation tests appears the shallowest, with errors being small as long as $\alpha<0.25$. This indicates that the fit of the exponential powers b${_1}$ and b${_2}$ is driven by the cyclic shear and biaxial tests. Among the three groups -- relaxation, cyclic shear and biaxial, the smallest errors are achieved in relaxation, as it can be seen in both Figs.~\ref{F:err_alpha} and~\ref{F:Model_fit}. 

Sensitivity analysis was conducted on the viscoelastic model and its parameters. Compared to the original values (figure~\ref{F:Model_fit}), the noise led to at most 4\% standard deviation and averages close to the reference values, as seen in figure~\ref{F:Noisy}.
This suggests that the model parameters are identifiable and robust to noise. A perturbation of $\pm$10\% in $\text{b}_\text{ff}$ and $\text{b}_\text{ss}$ leads to the model error increase of $\sim 1.5\%$, while $\pm10\%$ perturbation of $\alpha$ and $\text{b}_1$ leads to a significant error increase -- 3 to 11\%. 
Importantly, all parameters alter the objective function, showing the model parameters are observable.

\subsection{Model-based predictive response}

To examine the response of the proposed VE model, predictions of other tests not used for training were simulated, including biaxial stretch to different fiber/cross-fiber ratios (figure~\ref{F:prediction_biax}), frequency response (figure~\ref{F:VE_prediction_freq}), and different degrees of maximal cyclic shear (figure~\ref{F:shear_prediction}).
Applying biaxial stretch at different ratios, the proposed VE model shows excellent predictive behavior by capturing the inherent nonlinearity and hysteresis.
Importantly, it also demonstrates the inherent coupling of fiber/cross-fiber stretches seen in the data, whereby change in the stretch applied in one direction influences the stress in other directions.
Another predictive behavior of the VE model is shown in figure~\ref{F:VE_prediction_freq}, for biaxial stretch tests at 0.01, 0.033 and 0.1 Hz. 
The data show that changes in frequency of one order of magnitude yield a modest increase in hysteresis and peak stress amplitude (up to $\sim 19\%$).
Modest increase in the viscoelastic model is also observed ($\sim 41\%$), due to the fractional differential form.
While the predictive increase is higher, this model captures this modest growth and would likely perform better with multi-frequency data created at higher frequencies.

Lastly, the VE model behavior is investigated at lower cyclic shear levels. 
From the original study data~\cite{sommer2015biomechanical}, strain softening can be observed as the shear increases with fairly sustained hysteresis, see figure~\ref{F:shear_prediction} (left plot).
In contrast, the proposed VE model (center plot), predicts decreased hysteresis at lower shears and nested curves that do not exhibit strain softening.
However, the original data was acquired following a protocol of progressive increases in shear strain, with preconditioning cycles at each level.
This means that the sample was not preconditioned to the largest shear levels until later cycles.
To demonstrate this impact, the test was repeated using porcine myocardium (using the same experimental protocol) whereby the shear protocol was stepped from largest to smallest shear strain (right plot).
In this case, the passive tissue response resembles that predicted by the VE model.

\subsection{Comparison with other models}

To investigate the relative impact of the proposed VE model, results were compared with standard hyperelastic \HO and Costa models (see figure~\ref{F:Model_fit}).
As expected, both hyperelastic models exhibit a constant stress in relaxation and no hysteresis; however, both models do well at capturing the behavior of the data with errors of $ 24.17 $ and $ 24.95 \% $.
It can be seen that the \HO model performs better in describing the nonlinear trend of the biaxial test, with the errors in this group being $\sim 10\%$ smaller than for the Costa model. 
In contrast, the proposed model reduces the error metric by approximately a factor of 3 to $ 7.64 \% $, with the largest reduction in relaxation tests.
While these results are expected to improve (in part, due to the increase in parameters from 7-8 to 11), these tests provide an important benchmark for understanding the potential benefit of using a viscoelastic modeling approach.

Another important comparison is with the previous viscoelastic model published by G\"{u}ltekin \emph{et al.}~\cite{gultekin2016orthotropic} that fits the data utilized here.
This paper extended the \HO model using a nonlinear Maxwell approach~\cite{zhang2021compare} that relied on 18 parameters fit to each test separately. 
Because this model utilized a single relaxation time, the best fit to the relaxation data showed challenges capturing the entire decay spectrum. 
In contrast, the proposed VE model captures this decay spectrum while also fitting other datasets.
The model also shows difficulty reproducing the cyclic shear data, particularly compared to the model proposed.
In addition, the model fitting for the G\"{u}ltekin model requires determination of 14 nonlinear parameters, significantly increasing the cost of parameterization and analysis.

Examining the predictive responses, a key observation comes when comparing the biaxial data at varying levels of stretch, see figure~\ref{F:prediction_biax}.
In this figure, the classic \HO model shows an uncoupling of stretch along microstructural directions, whereby loading the fiber to the same stretch and varying the stretch across fibers produces nearly the same load.
This is in contrast to the Costa model and the experimental data, which show that these loads are not fully independent. 
In contrast, the VE model provides varying responses that qualitatively match the behavior of the data.
In our testing, adding fiber dispersion to the \HO model does show improvement; however, this tends to come at the expense of accuracy in other tests and adding these parameters did not substantially improve the model response.
As this model forms the basis of that presented in~\cite{gultekin2016orthotropic}, it is likely that similar challenges to biaxial prediction would be observed.

\subsection{Study limitations}

In this study, data from 14 tests were utilized to parameterize all models -- a challenge which is often not attempted within constitutive model studies.
While integration provides arguably a more complete result, the challenge comes that no tests stem from the same samples.
Hence, the inherent variability in tissues make the analysis challenging.
To circumvent this, all models were arbitrarily scaled based on testing groups (relaxation, cyclic shear, biaxial stretch).
This allowed models to capture the essential shape, without being heavily biased by the total amplitude.
An improvement to this study could be the use of new testing rigs~\cite{avazmohammadi2018integrated}, capable of providing a wide range of tests on a single sample.

\section{Conclusions}

In this paper, we present a nonlinear viscoelastic constitutive model for passive myocardium using a fractional approach.
The model is tested against human myocardial data across a range of testing protocols, demonstrating effective reproduction of experimental measurements as well as strong prediction of other tissue measures.
The model builds on the idea of cardiac viscoelasticity stemming from its hierarchical structure, yielding a spectrum of viscoelastic phenomena that space spatiotemporal scales.
The model is compared with classic hyperelastic models, demonstrating a significant improvement in fitting (mean error $ {\sim 7.65 \%} $ compared to $ {\sim25 \%} $), as well as in predictive response (particularly for variable biaxial stretch).
The model also is shown to exhibit a unique set of parameters that are observable and it is robust. 
The proposed VE model presents one of the first constitutive models aimed at capturing viscoelastic nonlinear response across multiple testing regimes, providing a platform for better understanding the biomechanics of myocardial tissue in health and disease.

\section{Declaration of competing interest}

The authors declare no competing interests.

\section{Acknowledgements}

DN acknowledges funding from the Engineering and Physical Sciences Research Council (EP/N011554/1) and the Engineering and Physical Sciences Research Council Healthcare Technology Challenge Award (EP/R003866/1), and support from the Wellcome Trust EPSRC Centre of Excellence in Medical Engineering (WT 088641/Z/09/Z) and the NIHR Biomedical Research Centre at Guy's and St.Thomas' NHS Foundation Trust and KCL. 
The views expressed are those of the authors and not necessarily those of the NHS, the NIHR, or the DoH.

\bibliographystyle{elsart-num-sort}
\bibliography{main}

\begin{thebibliography}{10}
\expandafter\ifx\csname url\endcsname\relax
  \def\url#1{\texttt{#1}}\fi
\expandafter\ifx\csname urlprefix\endcsname\relax\def\urlprefix{URL }\fi

\bibitem{adolfsson2003fractional}
K.~Adolfsson, M.~Enelund, Fractional derivative viscoelasticity at large
  deformations, Nonlinear dynamics 33~(3) (2003) 301--321.

\bibitem{avazmohammadi2018integrated}
R.~Avazmohammadi, D.~S. Li, T.~Leahy, E.~Shih, J.~S. Soares, J.~H. Gorman,
  R.~C. Gorman, M.~S. Sacks, An integrated inverse model-experimental approach
  to determine soft tissue three-dimensional constitutive parameters:
  Application to post-infarcted myocardium, Biomechanics and modeling in
  mechanobiology 17~(1) (2018) 31--53.

\bibitem{avazmohammadi2020vivo}
R.~Avazmohammadi, J.~S. Soares, D.~S. Li, T.~Eperjesi, J.~Pilla, R.~C. Gorman,
  M.~S. Sacks, On the in vivo systolic compressibility of left ventricular free
  wall myocardium in the normal and infarcted heart, Journal of Biomechanics
  107 (2020) 109767.

\bibitem{azeloglu2008heterogeneous}
E.~U. Azeloglu, M.~B. Albro, V.~A. Thimmappa, G.~A. Ateshian, K.~D. Costa,
  Heterogeneous transmural proteoglycan distribution provides a mechanism for
  regulating residual stresses in the aorta, American Journal of
  Physiology-Heart and Circulatory Physiology 294~(3) (2008) H1197--H1205.

\bibitem{benedicto2011structural}
H.~G. Benedicto, P.~P. Bombonato, G.~Macchiarelli, G.~Stifano, I.~M. Prado,
  Structural arrangement of the cardiac collagen fibers of healthy and diabetic
  dogs, Microscopy Research and Technique 74~(11) (2011) 1018--1023.

\bibitem{bishop1999regulation}
J.~E. Bishop, G.~Lindahl, Regulation of cardiovascular collagen synthesis by
  mechanical load, Cardiovascular Research 42~(1) (1999) 27--44.

\bibitem{blix1892lange}
M.~Blix, Die {L}{\"a}nge und die {S}pannung des {M}uskels, Skandinavisches
  Archiv f{\"u}r Physiologie 3~(1) (1892) 295--318.

\bibitem{blix1893lange}
M.~Blix, Die {L}{\"a}nge und die {S}pannung des {M}uskels, Skandinavisches
  Archiv f{\"u}r Physiologie 4~(1) (1893) 399--409.

\bibitem{blix1894lange}
M.~Blix, Die {L}{\"a}nge und die {S}pannung des {M}uskels, Skandinavisches
  Archiv f{\"u}r Physiologie 5~(1) (1894) 173--206.

\bibitem{Bonet1997}
J.~Bonet, R.~Wood, Nonlinear continuum mechanics for finite element analysis,
  Cambridge University Press, 1997.

\bibitem{cansiz2015orthotropic}
F.~B.~C. Cans{\i}z, H.~Dal, M.~Kaliske, An orthotropic viscoelastic material
  model for passive myocardium: theory and algorithmic treatment, Computer
  Methods in Biomechanics and Biomedical Engineering 18~(11) (2015) 1160--1172.

\bibitem{cavalcante2005mechanical}
F.~S. Cavalcante, S.~Ito, K.~Brewer, H.~Sakai, A.~M. Alencar, M.~P. Almeida,
  J.~S. Andrade~Jr, A.~Majumdar, E.~P. Ingenito, B.~Suki, Mechanical
  interactions between collagen and proteoglycans: implications for the
  stability of lung tissue, Journal of applied physiology 98~(2) (2005)
  672--679.

\bibitem{Chabiniok2016}
R.~Chabiniok, V.~Y. Wang, M.~Hadjicharalambous, L.~Asner, J.~Lee, M.~Sermesant,
  E.~Kuhl, A.~A. Young, P.~Moireau, M.~P. Nash, D.~Chapelle, D.~A. Nordsletten,
  Multiphysics and multiscale modelling, data{\textendash}model fusion and
  integration of organ physiology in the clinic: ventricular cardiac mechanics,
  Interface Focus 6~(2) (2016) 1--24.

\bibitem{cleutjens2002integration}
J.~P. Cleutjens, E.~E. Creemers, Integration of concepts: cardiac extracellular
  matrix remodeling after myocardial infarction, Journal of cardiac failure
  8~(6) (2002) S344--S348.

\bibitem{coleman1961foundations}
B.~D. Coleman, W.~Noll, Foundations of linear viscoelasticity, Reviews of
  Modern Physics 33~(2) (1961) 239.

\bibitem{costa2001modelling}
K.~D. Costa, J.~W. Holmes, A.~D. McCulloch, Modelling cardiac mechanical
  properties in three dimensions, Philosophical transactions of the Royal
  Society of London. Series A: Mathematical, physical and engineering sciences
  359~(1783) (2001) 1233--1250.

\bibitem{de1992internal}
P.~P. de~Tombe, H.~Ter~Keurs, An internal viscous element limits unloaded
  velocity of sarcomere shortening in rat myocardium., The Journal of
  Physiology 454 (1992) 619.

\bibitem{demer1983}
L.~L. Demer, F.~Yin, Passive biaxial mechanical properties of isolated canine
  myocardium., The Journal of Physiology 339~(1) (1983) 615--630.

\bibitem{dill2006continuum}
E.~H. Dill, Continuum Mechanics: Elasticity, Plasticity, Viscoelasticity, CRC
  press, 2006.

\bibitem{dokos2000triaxial}
S.~Dokos, I.~J. LeGrice, B.~H. Smaill, J.~Kar, A.~A. Young, A
  triaxial-measurement shear-test device for soft biological tissues, Journal
  of biomechanical engineering 122~(5) (2000) 471--478.

\bibitem{dokos2002}
S.~Dokos, B.~H. Smaill, A.~A. Young, I.~J. LeGrice, Shear properties of passive
  ventricular myocardium, American Journal of Physiology-Heart and Circulatory
  Physiology 283~(6) (2002) H2650--H2659.

\bibitem{fratzl2008collagen}
P.~Fratzl, Collagen: Structure and Mechanics, Springer Science \& Business
  Media, 2008.

\bibitem{fratzl1998fibrillar}
P.~Fratzl, K.~Misof, I.~Zizak, G.~Rapp, H.~Amenitsch, S.~Bernstorff, Fibrillar
  structure and mechanical properties of collagen, Journal of structural
  biology 122~(1) (1998) 119--122.

\bibitem{fung2013biomechanics}
Y.-C. Fung, Biomechanics: Mechanical Properties of Living Tissues, Springer
  Science \& Business Media, 2013.

\bibitem{guccione1991}
J.~M. Guccione, A.~D. McCulloch, L.~Waldman, Passive material properties of
  intact ventricular myocardium determined from a cylindrical model, Journal of
  Biomechanical Engineering 113~(1) (1991) 42--55.

\bibitem{gultekin2016orthotropic}
O.~G{\"u}ltekin, G.~Sommer, G.~A. Holzapfel, An orthotropic viscoelastic model
  for the passive myocardium: continuum basis and numerical treatment, Computer
  Methods in Biomechanics and Biomedical Engineering (2016) 1--18.

\bibitem{guo2007effect}
X.~Guo, Y.~Lanir, G.~S. Kassab, Effect of osmolarity on the zero-stress state
  and mechanical properties of aorta, American Journal of Physiology-Heart and
  Circulatory Physiology 293~(4) (2007) H2328--H2334.

\bibitem{Hadjicharalambous2015}
M.~Hadjicharalambous, R.~Chabiniok, L.~Asner, E.~Sammut, J.~Wong,
  G.~Carr-White, J.~Lee, R.~Razavi, N.~Smith, D.~Nordsletten, Analysis of
  passive cardiac constitutive laws for parameter estimation using {3D} tagged
  {MRI}, Biomechanics and Modeling in Mechanobiology 14~(4) (2015) 807--828.

\bibitem{hanley19993}
P.~J. Hanley, A.~A. Young, I.~J. LeGrice, S.~G. Edgar, D.~S. Loiselle,
  3-dimensional configuration of perimysial collagen fibres in rat cardiac
  muscle at resting and extended sarcomere lengths, The Journal of Physiology
  517~(3) (1999) 831--837.

\bibitem{hayes2013effect}
S.~Hayes, C.~S. Kamma-Lorger, C.~Boote, R.~D. Young, A.~J. Quantock, A.~Rost,
  Y.~Khatib, J.~Harris, N.~Yagi, N.~Terrill, et~al., The effect of
  riboflavin/uva collagen cross-linking therapy on the structure and
  hydrodynamic behaviour of the ungulate and rabbit corneal stroma, PloS One
  8~(1) (2013) e52860.

\bibitem{hill1921thermo}
A.~Hill, W.~Hartree, The thermo-elastic properties of muscle, Philosophical
  Transactions of the Royal Society of London. Series B, Containing Papers of a
  Biological Character 210 (1921) 153--173.

\bibitem{holm2019waves}
S.~Holm, Waves with Power-Law Attenuation, Springer, 2019.

\bibitem{holzapfel1996large}
G.~A. Holzapfel, On large strain viscoelasticity: continuum formulation and
  finite element applications to elastomeric structures, International Journal
  for Numerical Methods in Engineering 39~(22) (1996) 3903--3926.

\bibitem{holzapfel2000nonlinear}
G.~A. Holzapfel, Nonlinear Solid Mechanics: a Continuum Approach for
  Engineering science, John Wiley \& Sons Ltd, 2000.

\bibitem{holzapfel2001viscoelastic}
G.~A. Holzapfel, T.~C. Gasser, A viscoelastic model for fiber-reinforced
  composites at finite strains: Continuum basis, computational aspects and
  applications, Computer Methods in Applied Mechanics and Engineering 190~(34)
  (2001) 4379--4403.

\bibitem{holzapfel2009}
G.~A. Holzapfel, R.~W. Ogden, Constitutive modelling of passive myocardium: a
  structurally based framework for material characterization, Philosophical
  Transactions of the Royal Society of London A: Mathematical, Physical and
  Engineering Sciences 367~(1902) (2009) 3445--3475.

\bibitem{holzapfel1996new}
G.~A. Holzapfel, J.~C. Simo, A new viscoelastic constitutive model for
  continuous media at finite thermomechanical changes, International Journal of
  Solids and Structures 33~(20) (1996) 3019--3034.

\bibitem{hoskins2010normal}
A.~C. Hoskins, A.~Jacques, S.~C. Bardswell, W.~J. McKenna, V.~Tsang, C.~G. dos
  Remedios, E.~Ehler, K.~Adams, S.~Jalilzadeh, M.~Avkiran, et~al., Normal
  passive viscoelasticity but abnormal myofibrillar force generation in human
  hypertrophic cardiomyopathy, Journal of molecular and cellular cardiology
  49~(5) (2010) 737--745.

\bibitem{humphrey1989}
J.~Humphrey, F.~Yin, Biomechanical experiments on excised myocardium:
  theoretical considerations, Journal of Biomechanics 22~(4) (1989) 377--383.

\bibitem{humphrey1990a}
J.~D. Humphrey, R.~K. Strumpf, F.~C.~P. Yin, {Determination of a Constitutive
  Relation for Passive Myocardium: {I}. A New Functional Form}, Journal of
  Biomechanical Engineering 112~(3) (1990) 333--339.

\bibitem{humphrey1990b}
J.~D. Humphrey, R.~K. Strumpf, F.~C.~P. Yin, Determination of a constitutive
  relation for passive myocardium: {II}. parameter estimation, Journal of
  biomechanical engineering 112~(3) (1990) 340--346.

\bibitem{huyghe1991}
J.~M. Huyghe, D.~H. van Campen, T.~Arts, R.~M. Heethaar, The constitutive
  behaviour of passive heart muscle tissue: a quasi-linear viscoelastic
  formulation, Journal of Biomechanics 24~(9) (1991) 841--849.

\bibitem{irastorza2015mathematical}
R.~M. Irastorza, B.~Drouin, E.~Blangino, D.~Mantovani, Mathematical modeling of
  uniaxial mechanical properties of collagen gel scaffolds for vascular tissue
  engineering, The Scientific World Journal 2015.

\bibitem{leach1965effect}
J.~K. Leach, R.~S. Alexander, Effect of epinephrine on stress relaxation and
  distensibility of the isolated cat heart, American Journal of
  Physiology--Legacy Content 209~(5) (1965) 935--940.

\bibitem{legrice2005architecture}
I.~LeGrice, A.~Pope, B.~Smaill, The architecture of the heart: myocyte
  organization and the cardiac extracellular matrix, in: Interstitial Fibrosis
  in Heart Failure, Springer, 2005, pp. 3--21.

\bibitem{legrice1995}
I.~J. LeGrice, B.~H. Smaill, L.~Z. Chai, S.~G. Edgar, J.~B. Gavin, P.~J.
  Hunter, Laminar structure of the heart: ventricular myocyte arrangement and
  connective tissue architecture in the dog, American Journal of
  Physiology-Heart and Circulatory Physiology 269~(2) (1995) H571--H582.

\bibitem{levin1927viscous}
A.~Levin, J.~Wyman, The viscous elastic properties of muscle, Proceedings of
  the Royal Society of London. Series B, Containing Papers of a Biological
  Character 101~(709) (1927) 218--243.

\bibitem{lewinter1979time}
M.~M. Lewinter, R.~Engler, R.~S. Pavelec, Time-dependent shifts of the left
  ventricular diastolic filling relationship in conscious dogs, Circulation
  research 45~(5) (1979) 641--653.

\bibitem{li2020insights}
D.~S. Li, R.~Avazmohammadi, S.~S. Merchant, T.~Kawamura, E.~W. Hsu, J.~H.
  Gorman~III, R.~C. Gorman, M.~S. Sacks, Insights into the passive mechanical
  behavior of left ventricular myocardium using a robust constitutive model
  based on full 3d kinematics, Journal of the mechanical behavior of biomedical
  materials 103 (2020) 103508.

\bibitem{li2015modeling}
H.~Li, Y.~Zhang, Modeling of the viscoelastic behavior of collagen gel from
  dynamic oscillatory shear measurements, Biorheology 51~(6) (2015) 369--380.

\bibitem{lieleg2010structure}
O.~Lieleg, M.~M. Claessens, A.~R. Bausch, Structure and dynamics of
  cross-linked actin networks, Soft Matter 6~(2) (2010) 218--225.

\bibitem{lunkenheimer2006myocardium}
P.~P. Lunkenheimer, K.~Redmann, P.~Westermann, K.~Rothaus, C.~W. Cryer,
  P.~Niederer, R.~H. Anderson, The myocardium and its fibrous matrix working in
  concert as a spatially netted mesh: a critical review of the purported
  tertiary structure of the ventricular mass, European journal of
  cardio-thoracic surgery 29~(Supplement 1) (2006) S41--S49.

\bibitem{Macchiarelli2002}
G.~Macchiarelli, O.~Ohtani, S.~Nottola, T.~Stallone, A.~Camboni, I.~Prado,
  P.~Motta, A micro-anatomical model of the distribution of myocardial
  endomysial collagen, Histology and Histopathology 17 (2002) 699--706.

\bibitem{magin2006fractional}
R.~L. Magin, Fractional Calculus in Bioengineering, vol.~2, Begell House
  Redding, 2006.

\bibitem{mainardi2010fractional}
F.~Mainardi, Fractional Calculus and Waves in Linear Viscoelasticity: an
  Introduction to Mathematical Models, World Scientific, 2010.

\bibitem{meghezi2012effects}
S.~Meghezi, F.~Couet, P.~Chevallier, D.~Mantovani, Effects of a
  pseudophysiological environment on the elastic and viscoelastic properties of
  collagen gels, International journal of biomaterials 2012.

\bibitem{mori2012dynamic}
H.~Mori, K.~Shimizu, M.~Hara, Dynamic viscoelastic properties of collagen gels
  in the presence and absence of collagen fibrils, Materials Science and
  Engineering: C 32~(7) (2012) 2007--2016.

\bibitem{mori2013dynamic}
H.~Mori, K.~Shimizu, M.~Hara, Dynamic viscoelastic properties of collagen gels
  with high mechanical strength, Materials Science and Engineering: C 33~(6)
  (2013) 3230--3236.

\bibitem{Nolan2014}
D.~R. Nolan, A.~L. Gower, M.~Destrade, R.~W. Ogden, J.~P. McGarry, {A robust
  anisotropic hyperelastic formulation for the modelling of soft tissue},
  Journal of the Mechanical Behavior of Biomedical Materials 39 (2014) 48--60.

\bibitem{noll1958mathematical}
W.~Noll, A mathematical theory of the mechanical behavior of continuous media,
  Archive for rational Mechanics and Analysis 2~(1) (1958) 197--226.

\bibitem{obrien1966time}
L.~O'Brien, J.~Remington, Time course of pressure changes following quick
  stretch in tortoise ventricle, American Journal of Physiology--Legacy Content
  211~(3) (1966) 770--776.

\bibitem{ogden1997non}
R.~W. Ogden, Non-Linear Elastic Deformations, Dover Publications, 2013.

\bibitem{pinto1973mechanical}
J.~G. Pinto, Y.~Fung, Mechanical properties of the heart muscle in the passive
  state, Journal of Biomechanics 6~(6) (1973) 597--616.

\bibitem{pipkin1968non}
A.~Pipkin, T.~Rogers, A non-linear integral representation for viscoelastic
  behaviour, Journal of the Mechanics and Physics of Solids 16~(1) (1968)
  59--72.

\bibitem{podlubny1998fractional}
I.~Podlubny, Fractional Differential Equations: an Introduction to Fractional
  Derivatives, Fractional Differential Equations, to Methods of their Solution
  and some of their Applications, vol. 198, Academic press, 1998.

\bibitem{purslow1989strain}
P.~P. Purslow, Strain-induced reorientation of an intramuscular connective
  tissue network: implications for passive muscle elasticity, Journal of
  biomechanics 22~(1) (1989) 21--31.

\bibitem{purslow2008extracellular}
P.~P. Purslow, The extracellular matrix of skeletal and cardiac muscle, in:
  P.~Fratzl (ed.), Collagen: Structure and Function, Springer, 2008, pp.
  325--357.

\bibitem{puxkandl2002viscoelastic}
R.~Puxkandl, I.~Zizak, O.~Paris, J.~Keckes, W.~Tesch, S.~Bernstorff,
  P.~Purslow, P.~Fratzl, Viscoelastic properties of collagen: synchrotron
  radiation investigations and structural model, Philosophical Transactions of
  the Royal Society of London B: Biological Sciences 357~(1418) (2002)
  191--197.

\bibitem{reeve2014constitutive}
A.~M. Reeve, M.~P. Nash, A.~J. Taberner, P.~M. Nielsen, Constitutive relations
  for pressure-driven stiffening in poroelastic tissues, Journal of
  biomechanical engineering 136~(8) (2014) 081011.

\bibitem{schiessel1995generalized}
H.~Schiessel, R.~Metzler, A.~Blumen, T.~Nonnenmacher, Generalized viscoelastic
  models: their fractional equations with solutions, J. Phys. A: Math. Gen 28
  (1995) 6567--6584.

\bibitem{Schmid2006}
H.~Schmid, M.~Nash, A.~Young, P.~Hunter, Myocardial material parameter
  estimation -- a comparative study for simple shear, Journal of biomechanical
  engineering 128~(5) (2006) 742--750.

\bibitem{schmid2008}
H.~Schmid, P.~O'Callaghan, M.~Nash, W.~Lin, I.~LeGrice, B.~Smaill, A.~Young,
  P.~Hunter, Myocardial material parameter estimation, Biomechanics and
  modeling in mechanobiology 7~(3) (2008) 161--173.

\bibitem{sharma2014heart}
K.~Sharma, D.~A. Kass, Heart failure with preserved ejection fraction,
  Circulation research 115~(1) (2014) 79--96.

\bibitem{shen2011viscoelastic}
Z.~L. Shen, H.~Kahn, R.~Ballarini, S.~J. Eppell, Viscoelastic properties of
  isolated collagen fibrils, Biophysical journal 100~(12) (2011) 3008--3015.

\bibitem{simo1987fully}
J.~Simo, On a fully three-dimensional finite-strain viscoelastic damage model:
  formulation and computational aspects, Computer Methods in Applied Mechanics
  and Engineering 60~(2) (1987) 153--173.

\bibitem{sommer2015quantification}
G.~Sommer, D.~C. Haspinger, M.~Andr{\"a}, M.~Sacherer, C.~Viertler,
  P.~Regitnig, G.~A. Holzapfel, Quantification of shear deformations and
  corresponding stresses in the biaxially tested human myocardium, Annals of
  Thoracic Surgery of biomedical engineering (2015) 1--15.

\bibitem{sommer2015biomechanical}
G.~Sommer, A.~J. Schriefl, M.~Andr{\"a}, M.~Sacherer, C.~Viertler, H.~Wolinski,
  G.~A. Holzapfel, Biomechanical properties and microstructure of human
  ventricular myocardium, Acta biomaterialia 24 (2015) 172--192.

\bibitem{truesdell2004non}
C.~Truesdell, W.~Noll, The non-linear field theories of mechanics, Springer,
  2004.

\bibitem{tschoegl1989phenom}
N.~Tschoegl, The Phenomenological Theory of Linear Viscoelastic Behavior,
  vol.~1, Springer-Verlag Heidelberg, 1989.

\bibitem{walker1960potentiation}
S.~M. Walker, Potentiation and hysteresis induced by stretch and subsequent
  release of papillary muscle of the dog, American Journal of
  Physiology--Legacy Content 198~(3) (1960) 519--522.

\bibitem{weis2000myocardial}
S.~M. Weis, J.~L. Emery, K.~D. Becker, D.~J. McBride, J.~H. Omens, A.~D.
  McCulloch, Myocardial mechanics and collagen structure in the osteogenesis
  imperfecta murine (oim), Circulation Research 87~(8) (2000) 663--669.

\bibitem{wilhelm1975viscoelasticity}
F.~Wilhelm, Viscoelasticity, London: Springer-Verlag, t975, 1975.

\bibitem{wineman2009nonlinear}
A.~Wineman, Nonlinear viscoelastic solids: A review, Mathematics and Mechanics
  of Solids 14~(3) (2009) 300--366.

\bibitem{witzenburg2012mechanical}
C.~Witzenburg, R.~Raghupathy, S.~M. Kren, D.~A. Taylor, V.~H. Barocas,
  Mechanical changes in the rat right ventricle with decellularization, Journal
  of biomechanics 45~(5) (2012) 842--849.

\bibitem{Woods:1892}
R.~H. Woods, A few applications of a physical theorem to membranes in the human
  body in a state of tension, Transactions of the Royal Academy of Medicine in
  Ireland 10~(1) (1892) 417--427.

\bibitem{xu2013experimental}
B.~Xu, H.~Li, Y.~Zhang, An experimental and modeling study of the viscoelastic
  behavior of collagen gel, Journal of biomechanical engineering 135~(5) (2013)
  054501.

\bibitem{xu2013understanding}
B.~Xu, H.~Li, Y.~Zhang, Understanding the viscoelastic behavior of collagen
  matrices through relaxation time distribution spectrum, Biomatter 3~(3)
  (2013) e24651.

\bibitem{yang1991}
M.~Yang, L.~A. Taber, The possible role of poroelasticity in the apparent
  viscoelastic behavior of passive cardiac muscle, Journal of Biomechanics
  24~(7) (1991) 587--597.

\bibitem{Yin1987}
F.~C. Yin, R.~K. Strumpf, P.~H. Chew, S.~L. Zeger, {Quantification of the
  mechanical properties of noncontracting canine myocardium under simultaneous
  biaxial loading.}, Journal of Biomechanics 20~(6) (1987) 577--589.

\bibitem{zhang2021compare}
W.~Zhang, A.~Capilnasiu, D.~Nordsletten, Comparative analysis of nonlinear
  viscoelastic models across common biomechanical experiments, Journal of
  Elasticity (2021) 1--36.

\bibitem{zhang2020efficient}
W.~Zhang, A.~Capilnasiu, G.~Sommer, G.~A. Holzapfel, D.~A. Nordsletten, An
  efficient and accurate method for modeling nonlinear fractional viscoelastic
  biomaterials, Computer Methods in Applied Mechanics and Engineering 362
  (2020) 112834.

\bibitem{zile2004diastolic}
M.~R. Zile, C.~F. Baicu, W.~H. Gaasch, Diastolic heart failure -- abnormalities
  in active relaxation and passive stiffness of the left ventricle, New England
  Journal of Medicine 350~(19) (2004) 1953--1959.

\end{thebibliography}

\end{document}